\documentclass{pasj00}

\begin{document}
\SetRunningHead{Y. Fukazawa, K. Makishima, T. Ohashi}{ASCA Compilation of X-Ray Properties of Hot Gas in Elliptical Galaxies and Galaxy Clusters}
\Received{2004/04/12}
\Accepted{2004/09/09}

\title{ASCA Compilation of X-Ray Properties of Hot Gas in 
Elliptical Galaxies and Galaxy Clusters: Two Breaks in the
Temperature Dependences}

\author{Yasushi \textsc{Fukazawa}
}
\affil{Department of Physical Science, School of Science, 1-3-1
Kagamiyama, Higashi-Hiroshima, Hiroshima 739-8526}
\email{fukazawa@hirax6.hepl.hiroshima-u.ac.jp}

\author{Kazuo \textsc{Makishima}}
\affil{Department of Physics, School of Science, The University of Tokyo, 7-3-1 Hongo, Bunkyo-ku, Tokyo 113-0033}
\and
\author{Takaya {\sc Ohashi}}
\affil{Department of Physics, Faculty of Science, Tokyo 
Metroplitan University, Hachioji, Tokyo 192-03}

%

\KeyWords{galaxies: clusters: general ---   galaxies: elliptical and
lenticular, cD --- X-rays: galaxies: clusters ---  X-rays: ISM} 

\maketitle

\begin{abstract}

Utilizing ASCA archival data of about 300 objects of elliptical galaxies, 
groups, and clusters of galaxies,
we performed systematic measurements of the X-ray properties 
of hot gas in their systems, and compiled them in this paper.
The steepness of the luminosity--temperature (LT) relation, 
$L_{\rm X}\propto(kT)^{\alpha}$, in the range of $kT=$1.5--15 keV
is $\alpha=3.17\pm0.15$, consistent with previous measurements.
In the relation, we find two breaks at around ICM temperatures of 
1 keV and 4 keV: $\alpha=2.34\pm0.29$ above 4 keV, $3.74\pm0.32$ 
in 1.5--5 keV, and $4.03\pm1.07$ below 1.5 keV.
Such two breaks are also evident in the temperature and size relation.
The steepness in the LT relation at $kT>4$ keV is consistent 
with the scale-relation derived from the CDM model, 
indicating that the gravitational effect is dominant in richer clusters,
while poorer clusters suffer non-gravity effects.
The steep LT relation below 1 keV is almost attributed to X-ray faint
systems of elliptical galaxies and galaxy groups.
We found that the ICM mass within the scaling radius $R_{1500}$ 
follows the relation of $M_{\rm gas}\propto T^{2.33\pm0.07}$
from X-ray faint galaxies to rich clusters.
Therefore, we speculate that even such X-ray faint systems 
contain a large-scale hot gas, which is too faint
to detect.

\end{abstract}

\section{Introduction}

Elliptical galaxies, groups, and clusters of galaxies are known to
contain a large amount of gravitationally bound hot gas, which is
bright in the X-ray band.
Their gravitational mass ranges over 3--4 orders from 
$10^{12-13}M_{\odot}$ to $10^{16}M_{\odot}$, 
and the temperature of the hot gas represents the depth
of gravitational potential.
Therefore, the temperature dependences of the physical values of the 
hot gas are
interesting in terms of hierarchical structure formations of 
galaxies and clusters and the origin of hot gas.

The relationship between the X-ray luminosity, $L_{\rm X}$, and the 
temperature, $kT$, of hot gas has been well investigated 
[e.g., Matsumoto et al. (1997) and O'Sullivan et al. (2001) for
elliptical galaxies; 
Edge, Stewart (1991), David et al. (1993), Ponman et al. (1996), 
Markevitch (1998), Allen, Fabian (1998), and Xue, Wu (2000)
for clusters of galaxies],
since X-ray luminosities and temperatures can be easily measured.
Another merit of this relation is a weak dependence on the integration
radius, and thus it can be well defined.

On the other hand, the X-ray luminosity of hot gas depends on 
several properties, such as the hot gas temperature, mass, and 
spatial distribution.
Alternatively, 
the mass, $M_{\rm gas}$, of hot gas is more important and useful information 
for considering the characteristic of hot gas.
Arnaud and Raymond (1992) and David, Jones, and Forman (1995) 
derived the relation of the
ICM mass, $M_{\rm gas}$, and temperature, $kT$, although their sample was
small and the error of the ICM mass was large.
A detailed relation of $kT$--$M_{\rm gas}$ has recently been obtained 
by Mohr, Mathiesen, and Evrard (1999), 
who found a clear correlation of $M_{\rm
gas}\propto (kT)^{1.98}$ within the radius $R_{500}$, within 
which the averaged mass density is 500-times
as high as the cosmic critical density.
The information is still limited to rich clusters,
and a relation for lower temperature systems has not been obtained yet,
because of their X-ray faintness.
The ICM mass fraction is also an important quantity to constrain the
parameters of structure formation theories.
However, the correlation of the ICM mass with the cluster temperature is not
straightforwardly presented, because there is no clear integration
radius that is well justified.
Recently, radii within which the average mass density is higher by a
certain factor than the cosmic critical density are frequently applied.
This has theoretical meaning since structure-formation theories show
that clusters of galaxies can form against the cosmic expansion within
such defined radii; for example, $R_{180}$ (Navarro et al. 1995).
However, the radius $R_{\rm det}$, within which X-ray
emission was detected, seems to be much smaller than $R_{180}$ for 
elliptical galaxies and galaxy groups. 
Following Navarro, Frenk, and White (1995), 
$R_{180}$ is $\sim1$ Mpc for these objects
although the radius $R_{\rm det}$ is
at most 0.3--0.5 Mpc for galaxy groups (Mulchaey et al. 1996a) 
and 0.05--0.2 Mpc for elliptical galaxies (Matsushita 2001).
Since the radius $R_{\rm det}$ depends on the
sensitivity of instruments, systematic studies with the same instrument
are necessary to discuss it.

Concerning the origin of hot gas and the formation history of systems,
groups and poor clusters of galaxies are especially 
attractive because they are intermediate systems between individual 
galaxies and clusters of galaxies, as indicated by the following 
three phenomena.
First, the mass of hot gas in galaxy groups scatters widely
regardless of their similar stellar mass, $M_{\rm star}$ (Mulchaey et
al. 1996a).
The hot-gas-to-stellar mass ratio, $M_{\rm gas}/M_{\rm star}$,
ranges from 0.01 to 5, and the lower values are similar to those of elliptical
galaxies and the higher values are typical for clusters.
Second, non-gravity heating is strongly suggested for these 
low-temperature systems.
Ponman, Cannon, and Navarro (1999) discovered the entropy floor; 
the ICM entropy at the center of poor clusters exceeds the extrapolation
from the relation between the temperature and the entropy of rich clusters,
suggesting substantial non-gravitational effects on the ICM in poor
clusters.
The entropy in the ICM is important to study the thermodynamic history.
A preheating model, which considers significant energy input before
accreting of gas into the cluster potential, has been suggested 
to account for this (e.g., Tozzi, Norman 2001), and 
claimed by some observations (Finoguenov et al. 2001), 
although it is also inconsistent with the
observations (Ponman et al. 2003; Mushotzky et
al. 2003; Pratt, Arnaud 2003).
The contribution of non-gravity heating is also consistent with the picture 
suggested by studies of the metal abundances; significant fractions of metals
ejected from member galaxies have escaped from groups and poor clusters
of galaxies (Renzini et al. 1993; Fukazawa et al. 1996; Fukazawa
1997), indicating that galactic winds in the early galaxy formation
epoch give a vast amount of energy to the surrounding hot gas.
Third, Matsushita (1997, 2001) found that 
there are two types of elliptical galaxies: one exhibits a compact
X-ray emission where the mass of hot gas is several percent of stellar
mass, and the other shows an extended X-ray emission that is
represented by the double-$\beta$ model, and the mass of the hot gas is
comparable to that of the galaxy groups.
The latter type of elliptical galaxies has analogy with cD galaxy in
groups, whereas the galaxy concentration around them is ambiguous.
It is quite interesting why these two types exist.
The above three issues are thought to have some relation with each other, and
we can approach the unified picture by investigating the temperature
dependence of the hot-gas properties from elliptical galaxies to rich
clusters simultaneously.

Here, we performed systematic measurements of the hot-gas properties as a
function of the temperature, 
using ASCA (Tanaka et al. 1994) archival data without any selection criteria.
Since observations of most objects were not performed by
surveys, but proposed by many persons, the sample was not
complete and was not easy to derive the luminosity, temperature, 
or mass function.
The merits to utilize the ASCA data are as follows.
The ASCA GIS is the most sensitive to the diffuse faint X-ray emission 
at the cluster periphery among the previous launched missions,
thanks to its stable low background, long exposure, and wide field of view.
ASCA can measure a hard X-ray surface brightness that is less
contaminated by the central cool component 
(e.g., Fabian et al. 1994).
Moreover, the ASCA capability of resolving the emission lines
leads to accurate measurements of emission integral of hot gas.
Measurements with the same instrument and analysis procedure 
are free from
systematic calibration uncertainties among different instruments and analyses.
This work is an extension of the ASCA results on 40 nearby clusters
(Fukazawa 1997; Matsumoto et al. 2000).
Throughout this paper, we assume the Hubble constant to be
50$h_{50}$ km s$^{-1}$ Mpc$^{-1}$ and $q_0=0$, and the errors
represent the 90\% confidence range.

\section{Data Sample}
We utilized all of the ASCA data of elliptical galaxies and clusters.
Several clusters were observed more than once, and we chose the observation
with the longest exposure.
The total number of objects that we identified as elliptical galaxies and
clusters are 313, and are listed in table \ref{sample-lst}.
Some of objects cannot be utilized for deriving
various correlations, due to an unknown redshift (17), insignificant
detection (13), 
or contamination of the environmental X-ray emission 
(such as cluster emission around non-cD elliptical galaxies) (10).
In the following, we catalogued the physical properties of hot gas for all
of the objects that were significantly detected.

Among 17 objects without an available redshift, 
the Fe-K line was clearly detected from 3 clusters, and we obtained the
redshift by fitting the spectra with the redshift parameter free.
For the residual objects, we estimated them from the
luminosity--temperature relation of Edge, Stewart (1991), $L_{\rm
X}=1.0\times10^{43}(kT/{\rm keV})^{2.79}$ erg s$^{-1}$ (2--10 keV).
Here, the utilized temperature and the flux were derived by the following
analysis procedure.
The redshifts estimated here are summarized in table 3, as indicated 
by $\dagger$, and
were used to derive various physical quantities for these objects.
There are 13 objects whose X-ray emission was not significantly detected,
and we calculated the upper limits of the 0.5--2 keV flux within 6 arcmin from 
each cluster center, assuming a power-law spectrum with a photon index 
of 1.5.
In table \ref{upperlimit}, we summarized them.
Among 10 objects that are contaminated by the surrounding extended emission, 
we excluded NGC 4472, NGC 4406, NGC 4374, NGC 1404, NGC 499, and NGC 6034
in our correlation study, since the ambient cluster emission was too strong to
constrain their X-ray surface brightness profiles.
NGC 2865 was contaminated by the hard source, whose position was 
around $9^{\rm h}23^{\rm m}44^{\rm s}$ and $-23^{\circ}08'50''$ (J2000).
This object is extended in the Chandra archival image, and its
temperature and flux are 4--5 keV and $1.7\times10^{-12}$ erg s$^{-1}$
cm$^{-2}$ (0.5--10 keV), respectively.
Considering the luminosity--temperature relation of clusters, 
it is thought to be an uncatalogued distant cluster whose redshift is 0.2--0.3.
NGC 4291, CL 2236--04, and RX J1031.6--2607 are also contaminated by 
the hard X-ray point source.
Therefore, we excluded these four objects in our study.
For NGC 1316 and NGC 4649, the surrounding cluster emission is weak, and
thus we analyzed them, setting the cut-off radius by eye so that
the surrounding emission would not affect the result.

\begin{figure}[hptb]
\begin{minipage}[tbhn]{8cm}
\centerline{\includegraphics[width=8cm]{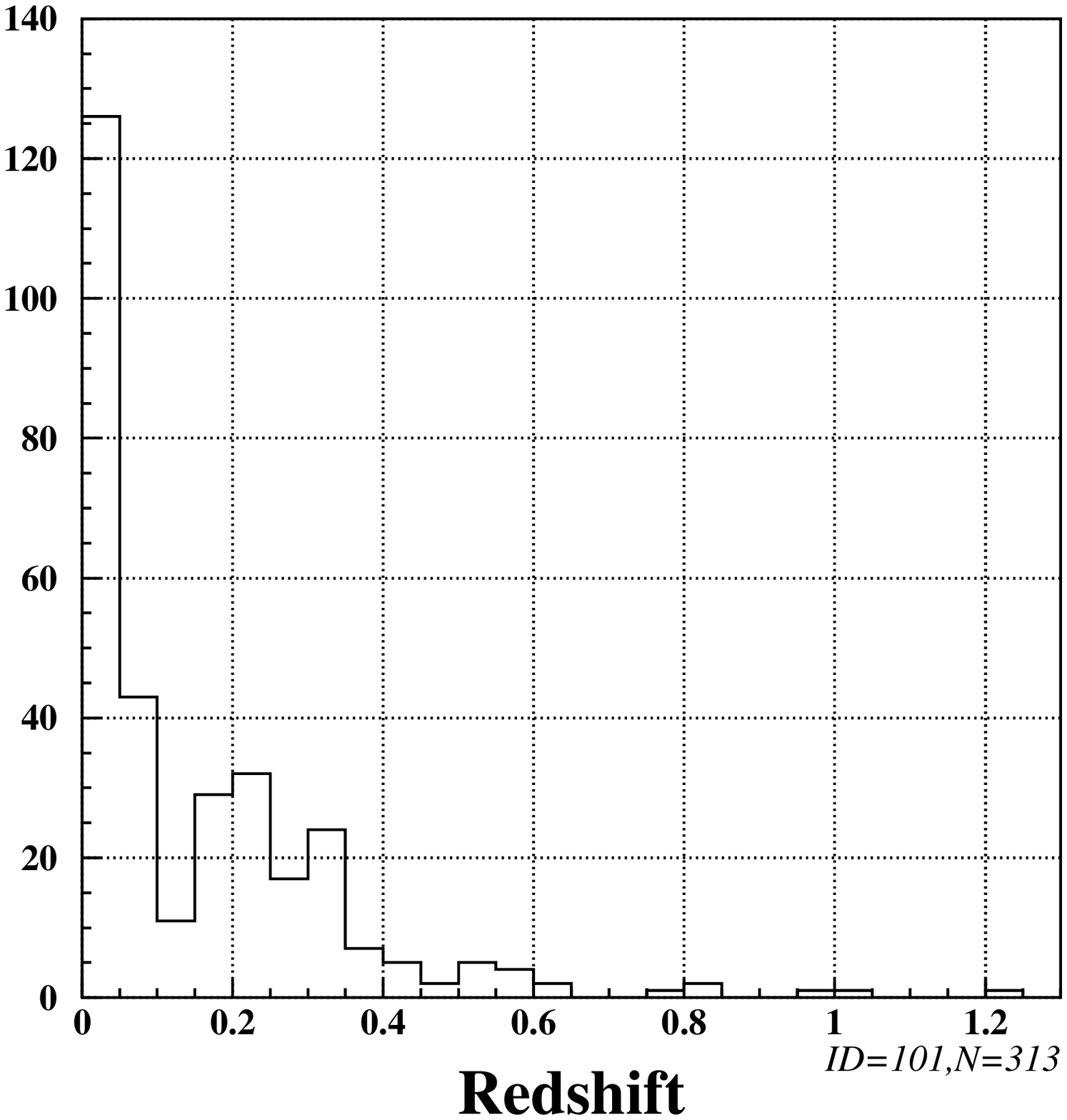}}
\caption{Redshift distribution of our sample objects.}
\label{zdist}
\end{minipage}\quad
\begin{minipage}[tbhn]{8cm}
\centerline{\includegraphics[width=8cm]{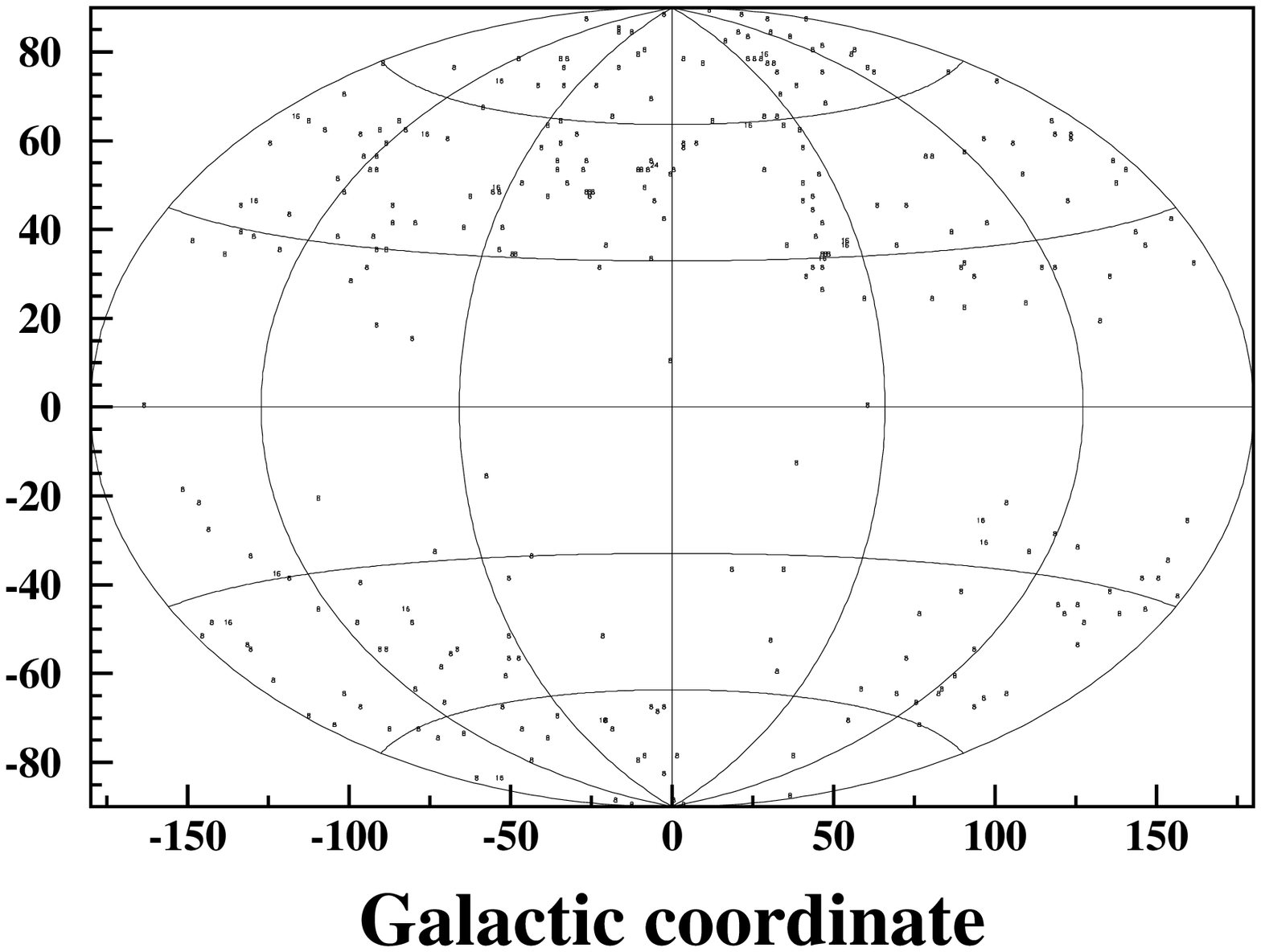}}
\caption{Distribution of our sample objects on the Galactic coordinate.}
\label{lbplot}
\end{minipage}
\end{figure}

As a result, we analyzed 292 objects, among which 
$\sim$50 elliptical galaxies and galaxy groups were included.
The redshift ranged from 0 to 0.8, mainly 0--0.4, as shown in figure 
\ref{zdist}.
Among them, we utilized 273 objects to derive various correlations.
In figure \ref{lbplot}, we show a plot of them on the Galactic coordinate.
Most objects locate at the high Galactic latitude, and our sample lacks
objects behind the Galactic plane.
However, we expect that this selection effect did not affect our results.

\begin{longtable}{ll}
\caption[]{Lists of clusters with upper limits of flux in the
observer frame at the 90\% confidence level.}
\label{upperlimit}
\hline
\hline
Name & $F_{\rm X}$ (0.5--2 keV) \\
 & (erg s$^{-1}$ cm$^{-2}$) \\
\hline
\endfirsthead
\endfoot
  \hline
\endlastfoot
NGC 5018 & $9.8\times10^{-14}$ \\ 
GHO 1322+3114 & $1.3\times10^{-13}$ \\ 
J 1888.16CL & $5.9\times10^{-14}$ \\ 
CL 0317+1521 & $4.5\times10^{-14}$ \\ 
MS 1512.4+3647 & $1.0\times10^{-12}$ \\ 
PRG 38 & $6.9\times10^{-14}$ \\ 
SCGG 205 & $6.9\times10^{-14}$ \\ 
RGH 101 & $9.1\times10^{-14}$ \\ 
3C 184 & $8.5\times10^{-14}$ \\ 
RX J1756.5+6512 & $1.6\times10^{-13}$ \\ 
3C 324 & $5.4\times10^{-14}$ \\ 
PDCS 01 & $2.8\times10^{-14}$ \\ 
MS 0147.8--3941 & $5.0\times10^{-14}$ \\ 
\hline
\end{longtable}

\section{Analysis}

\subsection{Data Reduction}

In the ASCA observations of the objects analyzed here,
the GIS (Gas Imaging Spectrometer; Ohashi et al. 1996; 
Makishima et al. 1996) data were all acquired in the normal 
PH mode, but the SIS (Solid-state Imaging Spectrometer) data were taken 
in various modes, such as the 4/2/1CCD Faint/Bright;
the 4CCD mode was usual at the early ASCA phase, while 2CCD or 1CCD was 
frequent at the later phase because of CCD radiation degradation.
Furthermore, the X-ray emission of objects is often over
a small field of view of the SIS.
For these reasons, we mainly analyzed the GIS data,
and the SIS was utilized only to constrain the spectral parameters of the 
low-temperature objects, such as elliptical galaxies and groups of galaxies.
The GIS data selections were performed on the condition of a minimum cut-off 
rigidity of 8 GeV c$^{-1}$ and a minimum elevation angle of 5$^{\circ}$ 
above the earth rim. 
Flare events have been known to occur a few times per day in the
GIS (Ishisaki 1996), and thus we excluded them.
For the SIS, we further imposed the condition of an elevation angle larger than 
25$^{\circ}$ above the day earth rim, and used events whose grade was 0, 2, 
3, or 4. 
We used the SIS data of only the FAINT mode 
after 1995, 1996, and 1997 for the 4, 2, and the 1CCD modes, respectively, 
and performed the RDD correction (Dotani et al. 1996).
In the analysis, we added all of the available data from different 
sensors, chips, modes, and pointings, separately for the GIS and the SIS 
after an appropriate gain correction.

At first, we constructed GIS images of each observation in the 0.6--7 keV, 
smoothed them with a Gaussian of $\sigma=0'.5$,
and searched for contaminating sources within 25$'$ of the detector
center as follows.
We searched pixels whose count rate was maximal, and marked it.
We stamped it as an X-ray source.
Next, we did the same thing, but by excluding
the regions within 3$'$ of the pixels that had already been marked.
If the marked pixel, 
except for the previous one, was not within 4$'$ of any marked
pixels, we considered it as an additional X-ray source.
We iterated the above procedure until the count rate of the newly marked
pixel became less than
the given value, which was the smaller one of either $2.5\times10^{-5}$c
s$^{-1}$ pix$^{-2}$ or 3 c pix$^{-2}$ ( 1 pix = $0'.25$ for thr GIS).
We thus typically excluded $2'.5$ from X-ray source candidates, or even
larger region for more extended or brighter sources for the GIS data.
The exclusion of the same region was applied for the SIS data.
After exclusion, we analyzed the GIS clean region on the detector
coordinate, as
defined in Ikebe (1995), which is defined so as to avoid a region 
where the background level is high or the calibration radio-isotope
contaminates significantly.

\subsection{Background Subtraction}

In order to constrain diffuse X-ray emission, especially at the cluster
periphery, we should subtract the GIS background as accurately as possible.
The background consists of the cosmic X-ray background (CXB)
and the intrinsic detector background (IDB).
In practice, for both image and spectral analyses,
we produced the background data set in the following way, which is almost 
the same as that described in Fukazawa et al. (2001), but a little different
because the ICM emission is sometimes extended beyond the GIS full field 
of view.

We first summed data of the ASCA Large Sky Surveys (Ueda et al. 1999),
conducted in 1993 December (AO-1 phase) and 1994 June (AO-2 phase)
over blank sky fields near the north ecliptic pole.
Then, after Ikebe (1995), Ishisaki (1996), and Ueda (1996),
we excluded regions in the GIS image
where the count rate exceeds that of the surrounding region 
by $\geq2.5\sigma$.
These regions correspond to faint sources with a
2--10 keV flux of $>8\times10^{-14}$ erg s$^{-1}$ cm$^{-2}$.
The total exposure time of the GIS background data amounts to 233 ksec,
ensuring negligible statistical errors.
However, the derived background data cannot be used immediately,
since the IDB level of the GIS has been gradually increasing by 2--3\% per 
year,
and it exhibits day-by-day fluctuations by 6--8\% in the standard deviation
(Ishisaki 1996; Ishisaki et al. 1997).

Taking into account these effects,
we estimated the IDB level of each pointing data individually, by assuming
that the IDB spectrum and its radial profile are both constant.
Specifically, we derived four GIS spectra,
denoted as $S(E)$, $B(E)$, $N(E)$, and $V(E)$;
the former three were obtained from the on-source data, the blank-sky
data, and the night-earth data, respectively.
The last one is a simulated spectrum of the ICM emission
that was estimated by a full XRT+GIS 
simulation based on the ICM temperature and spatial distribution, both
of which are first unknown.
For a background estimation, accurate information of the ICM is 
not necessary.
We then obtained it by ignoring the IDB time variation, by
utilizing the GIS background data set described above, which is correct
in a 0-th order approximation.
A detailed description about how to derive the ICM temperature and
spatial distribution is given in the next subsection.

\begin{figure}[hptb]
\centerline{\includegraphics[width=8cm]{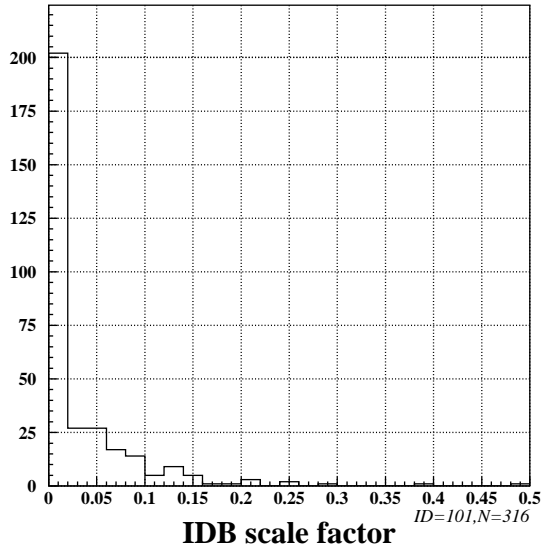}}
\caption{Distribution of the background correction factor, $f$, in the GIS
 data (see text in detail).}
\label{bgdist}
\end{figure}

These spectra were accumulated over the outer regions of the GIS field
in the 6--10 keV energy range,
to ensure that the CXB and $V(E)$ are relatively minor compared to the
IDB in $S(E)$ and $B(E)$.
Note that $N(E)$ consists solely of the IDB.
We next fitted $S(E)$ with a linear combination of $B(E)+fN(E)+V(E)$,
where $f$ is a free parameter,
and $fN(E)$ represents the IDB difference
between the two epochs when $S(E)$ and $B(E)$ were acquired.
Then, the fraction $f$ turned out to be around 0.00-0.30 for any
observations, as shown in figure \ref{bgdist}.
These values agree with the long-term increase of the GIS IDB
(Ishisaki et al. 1997).
By analyzing various ASCA data,
we also confirmed that this method can reproduce, within 5\%,
the GIS background spectra and radial profiles
acquired at any time over 1993--1999.
The background data obtained in this way were utilized
in the subsequent image and spectral analyses.

For SIS spectral analyses,
we subtracted the background in a conventional way,
utilizing the archival SIS background set.

\subsection{Analysis Procedure}

As described in the previous subsection, the background estimation
requires both spectral and imaging informations of objects.
Therefore, we first performed analyses by ignoring the gradual IDB increase
to set $f=0$, and
obtained the spectral and imaging parameters somewhat roughly, but they
were sufficiently
accurate to constrain the IDB level.
After estimating the IDB parameter, $f$, we again reanalyzed to
constrain the spectral and imaging parameters more accurately.

In the spectral and imaging analyses, 
we first determined two radial extents for each object, 
$R_{\rm 1E6}$ and $R_{\rm 3E6}$, 
within which the GIS count rate was
higher than $1\times10^{-6}$ and $3\times10^{-6}$
c s$^{-1}$ pix$^{-2}$, respectively, in the 0.9--7 keV.
The GIS2+3 azimuthally averaged radial count rate profile after the
background subtraction was used to determine them.

\subsection{Spectral Fittings}

\begin{figure}[hptb]
\centerline{\includegraphics[width=9.5cm]{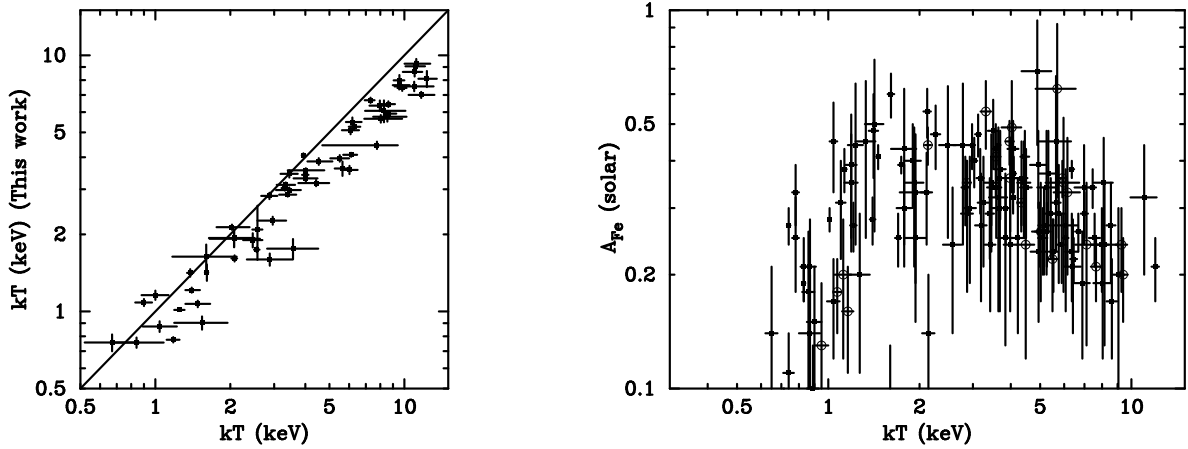}}
\caption{Comparison of temperatures between our results and those of Sanderson et al. (2003).}
\label{tcmp}
\end{figure}

We here performed spectral fittings with the IDB factor $f$ free, and 
obtained the ICM temperature.
The choice of integration radius for the spectral analysis followed the 
rule given in table \ref{r-ana}, which
was determined so as to avoid the central cool region.
For objects whose spectrum is statistically poor, we set the inner
radius to be smaller so as to include the bright center region.
The ancillary response file (arf file) was calculated by averaging arf
files at various positions, prepared by a step of $2'.5$ from the XRT
optical axis.
The spectral model applied here was the MEKAL model (Liedahl et al. 1995)
multiplied by the photoelectric absorption.
The column density was fixed to the Galactic value (Stark et al. 1992).
In measurements of the temperature, the metal abundance was assumed to be 
0.25 solar abundance, which refers 
to the solar photospheric values in Anders and Grevesse (1989), in order
to minimize the error, especially for faint objects.
Basically, we utilized only the GIS data, since the SIS covers only
the inner region of clusters due to its small field of view, or the
SIS data at a later period often exhibit a severe RDD that
changes a real spectral shape and cannot be completely corrected.
On the other hand, for objects with $kT<2$ keV, we performed 
simultaneous GIS+SIS spectral fittings because the SIS can well constrain the
low-temperature component.
In addition, we included a bremsstrahlung model for objects with
$kT<1.5$ keV, to
reproduce the emission from low-mass X-ray binaries in elliptical
galaxies (Matsushita et al. 1994), or the unresolved excess hard X-ray
emission from groups of galaxies (Fukazawa et al. 2001).
Spectral fittings were performed in the 0.85--9.0 keV region for the GIS and
0.5--8.0 keV for the SIS.
The thus-obtained results of the spectral fitting after considering the IDB 
increase are summarized in table \ref{specfit-lst}.
Besides 12 distant clusters and 2 poor clusters,
the temperatures are well constrained with a $<20$\% accuracy.
In figure \ref{tcmp}, we compare the temperature of our results with that in
Sanderson et al. (2003).
Our temperatures are systematically lower than those of Sanderson et
al. (2003).
Since Sanderson et al. (2003) determined the emission-weighted
temperature by extrapolating the negative radial gradient of the temperature
toward the central cooling region, the temperature is thought to become 
higher due to large emissivity at the center.

\begin{figure}[hptb]
\begin{minipage}[tbhn]{8cm}
\centerline{\includegraphics[width=9.5cm]{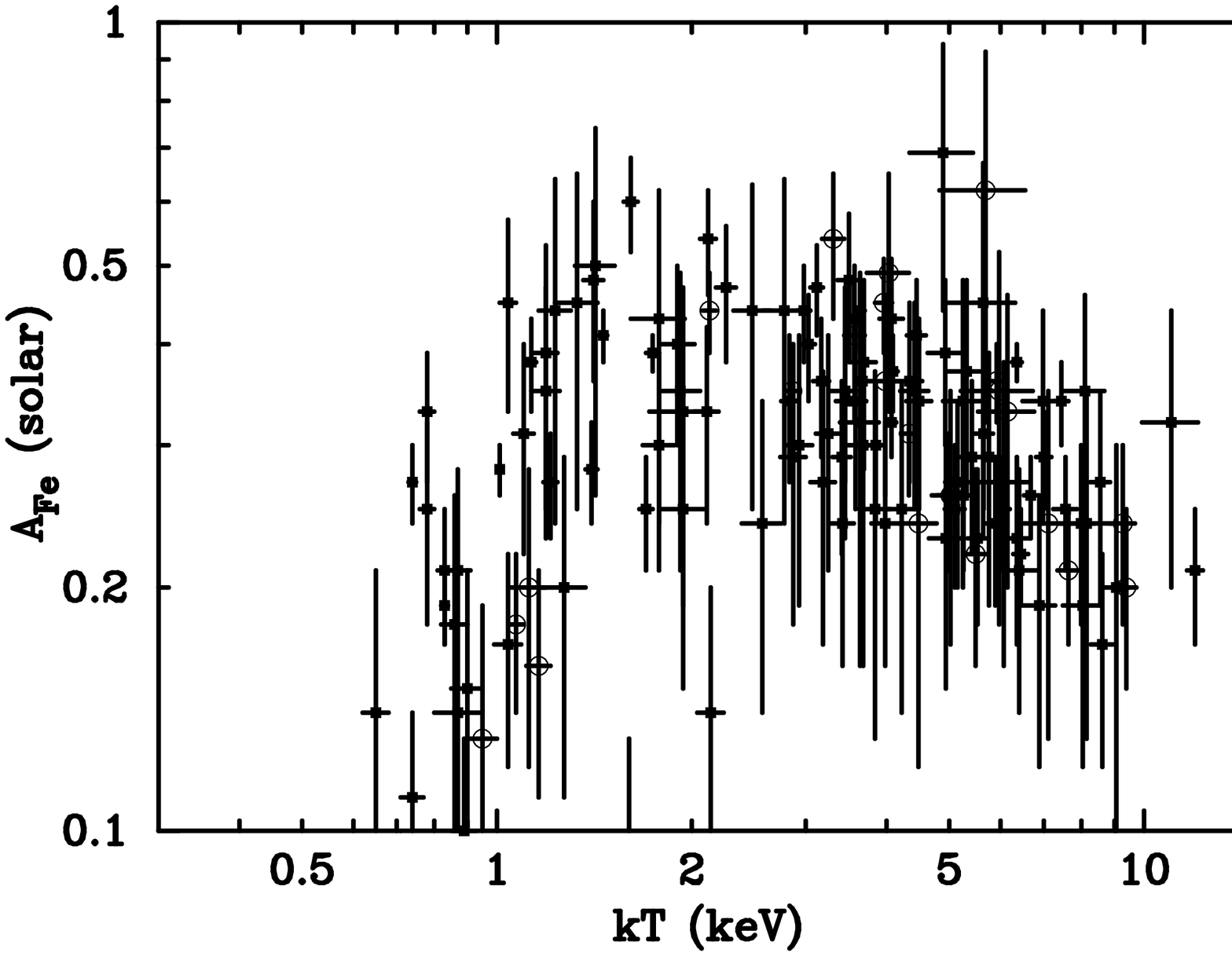}}
\end{minipage}\quad
\begin{minipage}[tbhn]{8cm}
\centerline{\includegraphics[width=9.5cm]{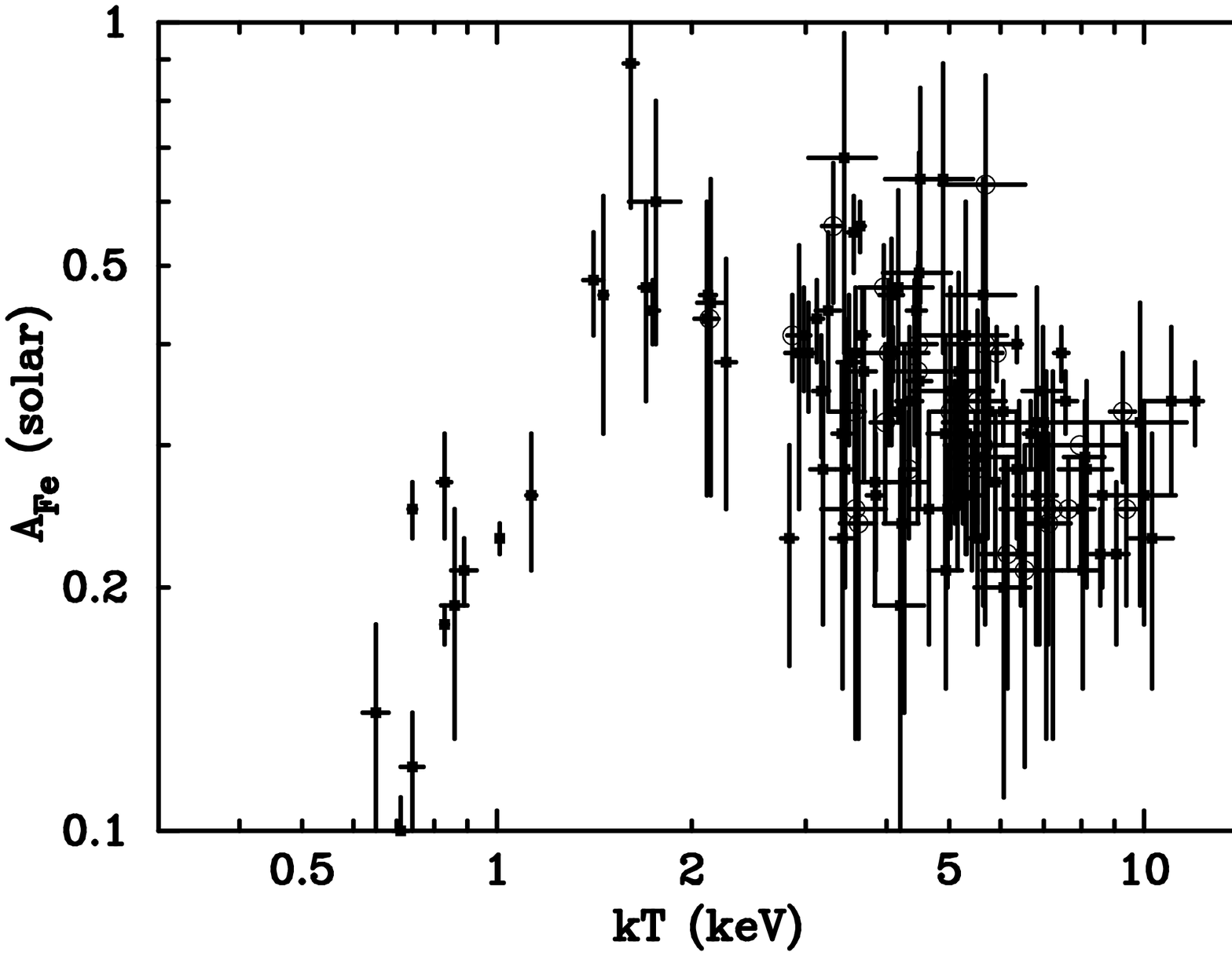}}
\end{minipage}
\caption{Fe abundance against the object temperature. The left panel
 plots the results for the spectral fitting in the default analysis
 region. The right panel plots for the whole cluster region.}
\label{tfe}
\end{figure}

Next, we performed a spectral fitting with the metal abundance free to
constrain it.
As a result, it was constrained with a $<30$\% accuracy for about 27 
objects.
In figure \ref{tfe} (left), 
the obtained metal abundances are plotted against the temperature.
Here, we do not plot objects whose error of the metal abundance is $>50$\%.
Above $kT>2$ keV, the metal abundance seems to negatively correlate with
the temperature.
This trend can be explained by the effect of the high-metallicity region 
at the bright cluster core, as claimed by Fukazawa et al. (2000), who showed
that the Fe abundance at the outer region is almost constant to be 0.2--0.3
solar for any temperature (Fukazawa et al. 1998).
Several reports on the metallicity increment at the center of clusters
support this explanation (Ezawa et al. 1997; Fukazawa et al. 2000;
De Grandi et al 2001).
In the choice of inner radius in this analysis, we make it
as small as possible so as to accumulate many photons.
In order to look at this effect more effectively, we plot the
emission-weighted metal abundance in figure \ref{tfe} (right), 
which was obtained from the spectra over the cluster.
The negative correlation of the metal abundance with the temperature becomes
prominent.
The high-redshift clusters exhibit a similar trend, in good
agreement with previous studies (Mushotzky, Loewenstein 1997; 
Matsumoto et al. 2000).
Accordingly, 
when we study the metal abundance of distant clusters that cannot be
resolved spatially, we must pay attention to the effect of high
metallicity at the center region.
Below $kT<1$ keV, the correlation disappears, and many objects exhibit
a low metal abundance.
This might be due to problems concerning the plasma modeling and
multi-temperature effects (Fukazawa et al. 1996; Buote 2000; Matsushita
et al. 2003).

\begin{longtable}{ll}
\caption[]{Definition of radial and energy boundaries for spectral and imaging analyses.}
\label{r-ana}
\endfirsthead
\endfoot
\hline
\hline
\multicolumn{2}{c}{Radial boundary for spectral analysis} \\
\hline
$R_{\rm 3E6}>10'$ & $3'$--$10'$ \\
$6'.75<R_{\rm 3E6}<=10'$ & $3'$--$R_{3E6}'$ \\
$5'<R_{\rm 3E6}<=6'.75$ & $2'$--$R_{3E6}'$ \\
$R_{\rm 3E6}<=5'$ & $1'$--$5'$ \\
\hline
\multicolumn{2}{c}{Boundary for imaging analysis} \\
\hline
\multicolumn{2}{c}{Radius} \\
$R_{\rm 1E6}>=15'$ & $0'$--$22'.5$ \\
$10'<R_{\rm 1E6}<15'$ & $0'$--$1.5r_0'$ \\
$R_{\rm 1E6}<=10'$ & $0'$--$10'$ \\
\hline
\multicolumn{2}{c}{Energy (keV)} \\
$kT<=10$(keV) & $1.0+0.767\frac{kT({\rm keV})-1}{2}$  (lower energy limit) \\
& $3.54+3.54\frac{kT({\rm keV})-1}{2}$ (higher energy limit) \\
$kT>10$(keV) & 4.45--19.5 \\
\hline
\end{longtable}

\subsection{Imaging Analysis}

After determining the ICM temperature, 
the azimuthally averaged radial count rate profile was fitted with the 
single-$\beta$ model.
The column density and metal abundance were fixed to the same value as
in the spectral analysis of the temperature.
We assumed isothermality, where
the ICM temperature is that obtained by spectral
fitting, and simulate the incident energy spectra with the 
plasma emission model by multiplying the photoelectric absorption.
Here, we applied the Raymond--Smith plasma model (Raymond, Smith 1977) 
rather than the MEKAL model in order to save time.
The integration radii and energy band for imaging analyses followed the
rule in table 2.
The azimuthally averaged radial profiles of the X-ray counts were constructed
by summing the GIS2 and GIS3 data.
The center of the X-ray emission was set to the position where the X-ray
brightness became maximal.
In order to model an X-ray surface brightness,
we made use of a software system, called an ASCA simulator 
(Ikebe 1995), in which 
we produced the XRT+GIS angular response by using the Cyg X-1 images 
actually observed with the GIS at 6 positions in the field of view 
(Takahashi et al. 1995). 
The actual radial profiles were compared with the simulated ones through a
$\chi^2$ evaluation, where we assigned 5\% systematic errors to the 
background and the simulated profile, according to Ikebe (1995).

We represented the ICM density distribution with a standard 
single-$\beta$ model (Cavaliere, Fusco-Femiano 1976) for all objects. 
However, an additional emission component is often required for poor
clusters in the single $\beta$ model fitting.
In such a case, we also applied a double-$\beta$ model
that succeeded the representation of the X-ray surface brightness for some
groups, rather than the simple-$\beta$ model 
(Ikebe et al. 1996; Mulchaey, Zabludoff 1998).
The double-$\beta$ model consists of a small-scale central component and
a large-scale component.
The X-ray surface brightness was expressed within a specified 
radius, $R_{\rm limit}$,
outside of which we assumed the ICM density to be zero.
We determined $R_{\rm limit}$ by increasing it until the $\chi^2$ did
not improve significantly.
The surface brightness profile was thus calculated as
\[\sum_{i=1}^k\int n_i(r)^2\Lambda\left(T\right)dl\],
where $r$ is the 3-dimensional radius, $l$ is toward the line-of-sight 
direction, 
and $\Lambda\left(T\right)$ is a 
cooling function calculated with the Raymond--Smith model by assuming 
0.25 solar abundance;
$k$ is 1 or 2 for a single or double-$\beta$ model, respectively, and
$i$ takes a value of 1 or 2 for a large-scale or small-scale component
in the double-$\beta$ model.
The square average of the electron density distribution, $n_i^2(r)$, 
is expressed as
 \[n_i^2(r)=n_{0,i}^2\left[1+\left(\frac{r}{a_i}\right)^{2}\right]^{-3\beta_i}\qquad\mbox{()i=1,2)},\]
where $n_{0,i}^2$ is a central value and $a_i$ is the core
radius.
For the double-$\beta$ model, the electron density is expressed as $\sqrt{n_1^2(r)+n_2^2(r)}$.
Since the spatial resolution of the XRT + GIS is poor to constrain $a_2$
and $\beta_2$ of the small-scale component in the double-$\beta$ model, 
we fixed them to be 
$a_2=10 h_{50}^{-1}$ kpc and $\beta_2=0.7$, which are averaged
values of the small-scale components in Mulchaey and Zabludoff (1998).
Therefore, the free parameters are $R_{\rm limit}$, $n_{0,1}$, $a_1$, 
and $\beta_1$ in the
single-$\beta$ model fitting, and $n_{0,2}$ becomes additional in the
double-$\beta$ model fitting.

Instead of building an automatic algorithm of $\chi^2$ minimization, we
generated a model profile at each of the $8\times8$ grid points in
$\beta_1$--$a_1$ space, and calculated $\chi^2$ by
adjusting the model normalization so as to archive the best fit between
the data and the simulation.
After this, we performed the same procedure by narrowing the
$\beta_1$--$a_1$ space around the above best-fit value.
From the normalization, we calculated the central value, $n_{0,1}^2$,
by assuming the ratio of electrons to ions to be 1.18 and the luminosity 
distance to the object to be 
\[
D=\frac{zq_0+\left(q_0-1\right)\left(\sqrt{2zq_0+1}-1\right)}{H_0q_0^2}c,\]
where $q_0=0.5$ and $H_0=50h_{50}$ km s$^{-1}$ Mpc$^{-1}$.
For elliptical galaxies and some galaxy groups, the hard component is
significant in the spectra.
Although we excluded the hard X-ray band in the image analysis, 
the contamination to the count rate must be taken into account.
For this purpose, we estimated the photon count fraction of the thermal
component and considered it in converting the surface brightness into
the flux or $n_{0,1}^2$.

\begin{figure}[hptb]
\begin{minipage}[tbhn]{8cm}
\centerline{\includegraphics[width=9.5cm]{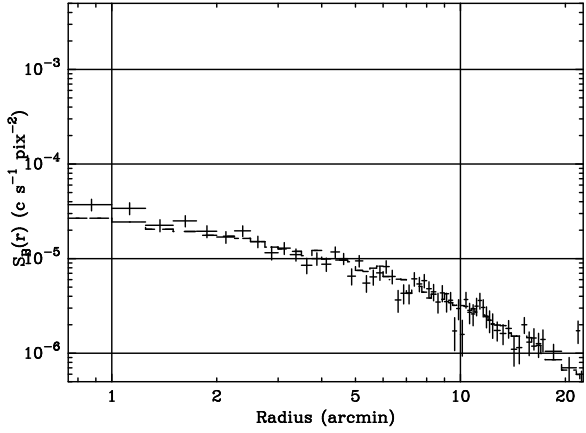}}
\end{minipage}\quad
\begin{minipage}[tbhn]{8cm}
\centerline{\includegraphics[width=9.5cm]{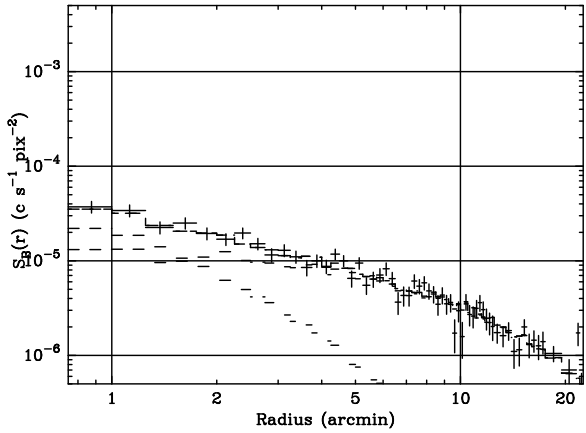}}
\end{minipage}
\caption{Examples of fittings of the radial count rate profile. The
 solid line represents the best-fit model and the cross points are the
 GIS data of the NGC 1600 group. (Left) single-$\beta$ model, (right) double-$\beta$
 model.}
\label{imfit}
\end{figure}

Examples of the best fittings are shown in figure
\ref{imfit}, where the best-fit model is plotted as a solid line, 
superposed on the actual count-rate profile.
The best-fit parameters and the lower limits of $R_{\rm limit}$ are 
tabulated in table \ref{b1fit-lst}.
The single-$\beta$ model fitting was successful in most cases with a reduced 
$\chi^2$ of 0.8--1.2.
The parameter errors for bright objects correspond to the parameter step,
and thus the true errors may be smaller.
The radius $R_{\rm limit}$ often corresponds to the radius of the GIS field
of view of $\sim20'$.
$\beta_1$ and $a_1$ do not change beyond their parameter search
step, as long as $R_{\rm limit}$ is within the confidence range.
There are 57 objects whose X-ray surface brightness profile is well
represented by the double-$\beta$ model, rather than the single-$\beta$ model.
In the following, we adopt the result of the double-$\beta$ model for these
57 objects, which are summarized in table \ref{b2fit-lst}.

\begin{figure}[hptb]
\centerline{\includegraphics[width=9.5cm]{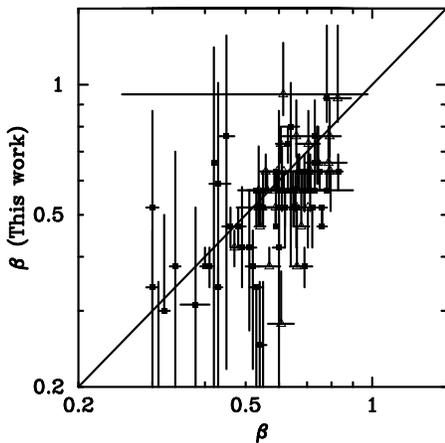}}
\caption{Comparison of $\beta$ of our results with those in Mohr, 
Mathiesen, and Evrard (1999) and Sanderson et al. (2003).}
\label{bcmp}
\end{figure}

From the tables, it is found that we can determine the core radius, $a_1$,
down to even
$\sim1'$ with the ASCA data in this method.
We compare the best-fit $\beta$ with that in Mohr, Mathiesen, and Evrard 
(1999) and 
Sanderson et al. (2003) in figure \ref{bcmp}.
It can be seen that our results for $\beta$ are in agreement with 
previous ones obtained in the soft X-ray band.
However, 
our $\beta$ and core radii are systematically smaller than those of
previous studies, by 5--10\% and 10--50\%, respectively.
Since these two quantities correlate with each other in the fitting, this could
be an instrumental artifact caused by, e.g., uncertainties in the XRT-PSF,
differences in the fields of view, and so on.
On the other hand, this might indicate that the X-ray surface brightness 
in the hard X-ray band is different from that in the soft X-ray band.
A more detailed discussion on the X-ray surface brightness is beyond the
scope of this paper.

\subsection{Temperature Determination under Full Consideration of the 
XRT-PSF Effect}

\begin{figure}[hptb]
\begin{minipage}[tbhn]{8cm}
\centerline{\includegraphics[width=8cm]{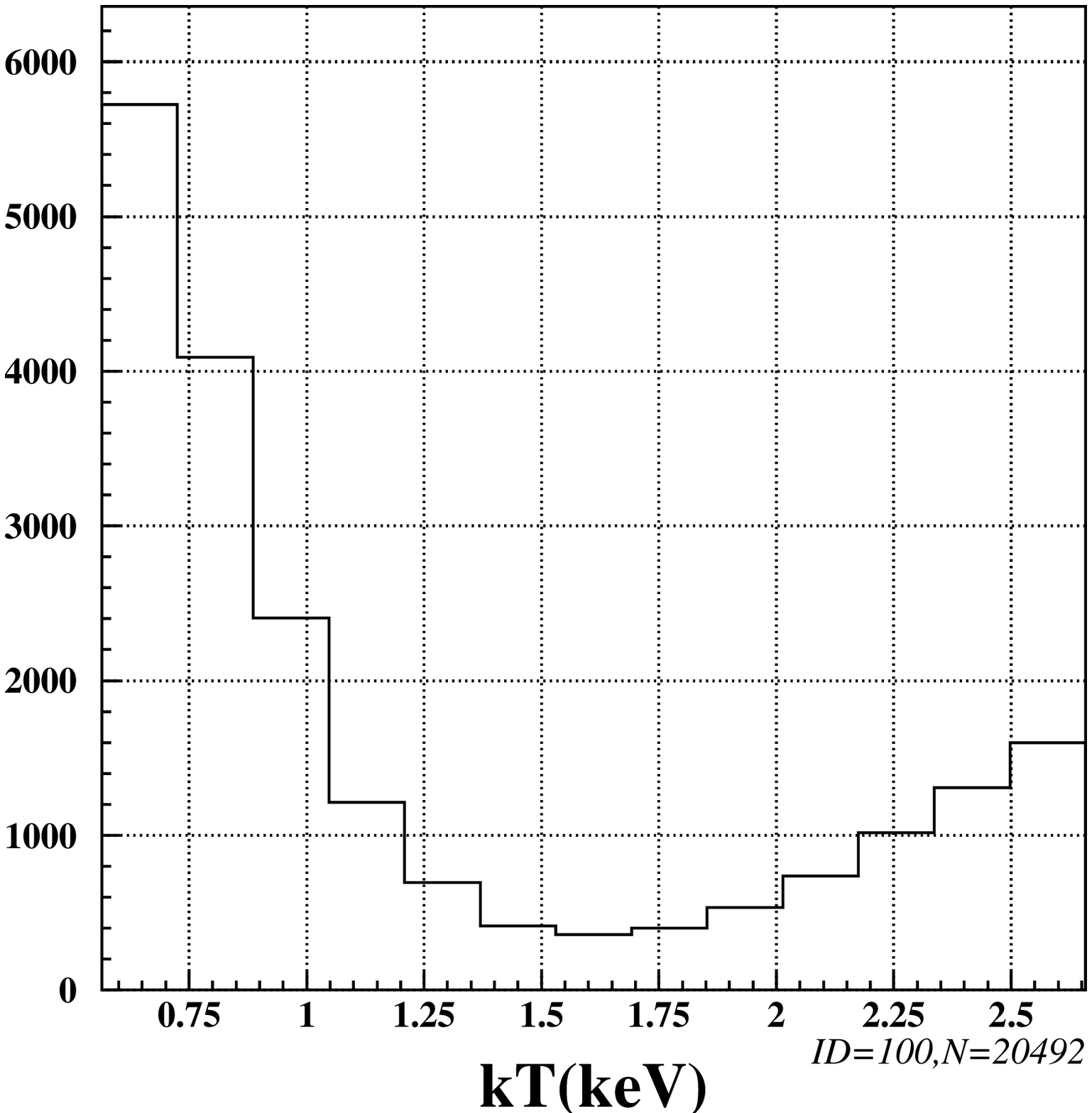}}
\caption{Example of a $\chi^2$ scan in the spectral fitting,
 considering the XRT-PSF effect for MKW 4. The vertical axis represents
 the $\chi^2$ value.}
\label{tfitchi}
\end{minipage}\quad
\begin{minipage}[tbhn]{8cm}
\centerline{\includegraphics[width=9.5cm]{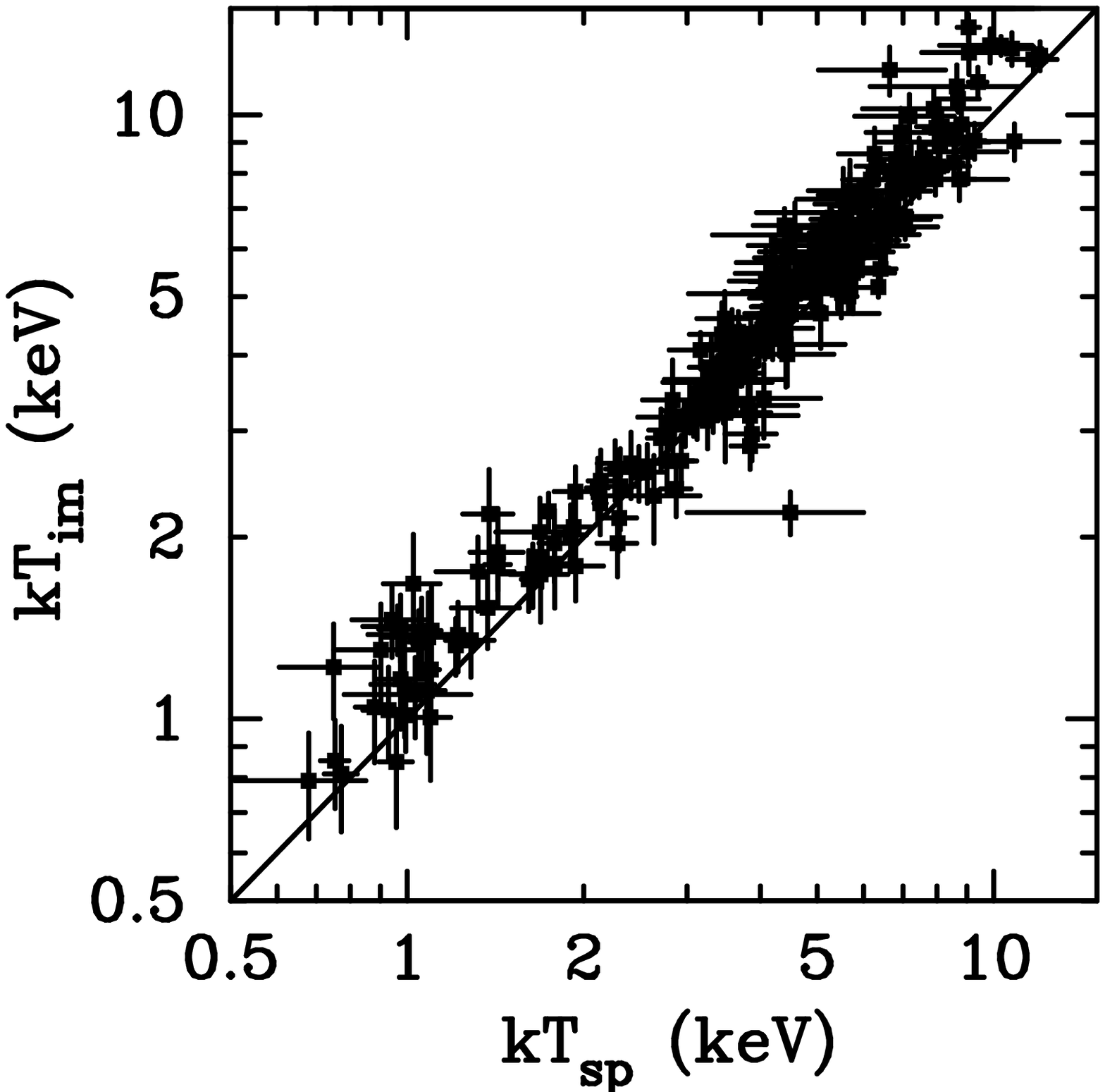}}
\caption{Comparison of the temperature obtained here $T_{\rm im}$ with that 
obtained in a simple spectral fitting ($T_{\rm sp}$).}
\label{tfitcmp}
\end{minipage}
\end{figure}

After obtaining imaging parameters, we further fitted the ICM
temperature by considering the XRT scattering effect, 
since the temperature obtained by simple spectral fittings 
might still be affected by XRT-PSF effects for high-temperature clusters
(Takahashi et al. 1995) that depend on the X-ray emission
distribution.
In this study, we excluded elliptical galaxies surrounded
by the intracluter medium.
Since the vignetting function strongly depends on the photon energy,
we must deal with not the emission-weighted single
spectrum, but the spectral sets consisting of spectra at several radii
for determining the temperature.
We obtained two or three spectra at an appropriate annulus for both the
data and simulation, and calculated the $\chi^2$ value of the difference 
between the data and the simulated spectral set by adjusting the normalization.
In preparing the simulated data, we modeled the ICM density distribution 
by the isothermal single-$\beta$ model with the
obtained single-$\beta$ model parameters.
We scanned the temperature from $0.4T_{\rm s}$ to $1.6T_{\rm s}$ by
steps of $0.2T_{\rm s}$, 
where $T_{\rm s}$ is the temperature obtained in the simple 
spectral fitting, and calculated the chi-square value $\chi^2$ 
for each temperature.
Then, the best-fit temperature, $T_i$, was determined by fitting the 
$T$--$\chi^2$ curve with the 2nd polynomial function, as
shown in figure \ref{tfitchi}.
We compared the results 
with the temperature obtained by the simple spectral fitting, 
in figure \ref{tfitcmp}.
The temperature differs by at most 10\% between the two estimations; 
hereafter, we thus adopt the temperature obtained by simple
spectral fittings.

\subsection{Flux, Luminosity, and ICM Mass}

\begin{figure}[hptb]
\begin{minipage}[tbhn]{8cm}
\centerline{\includegraphics[width=9.5cm]{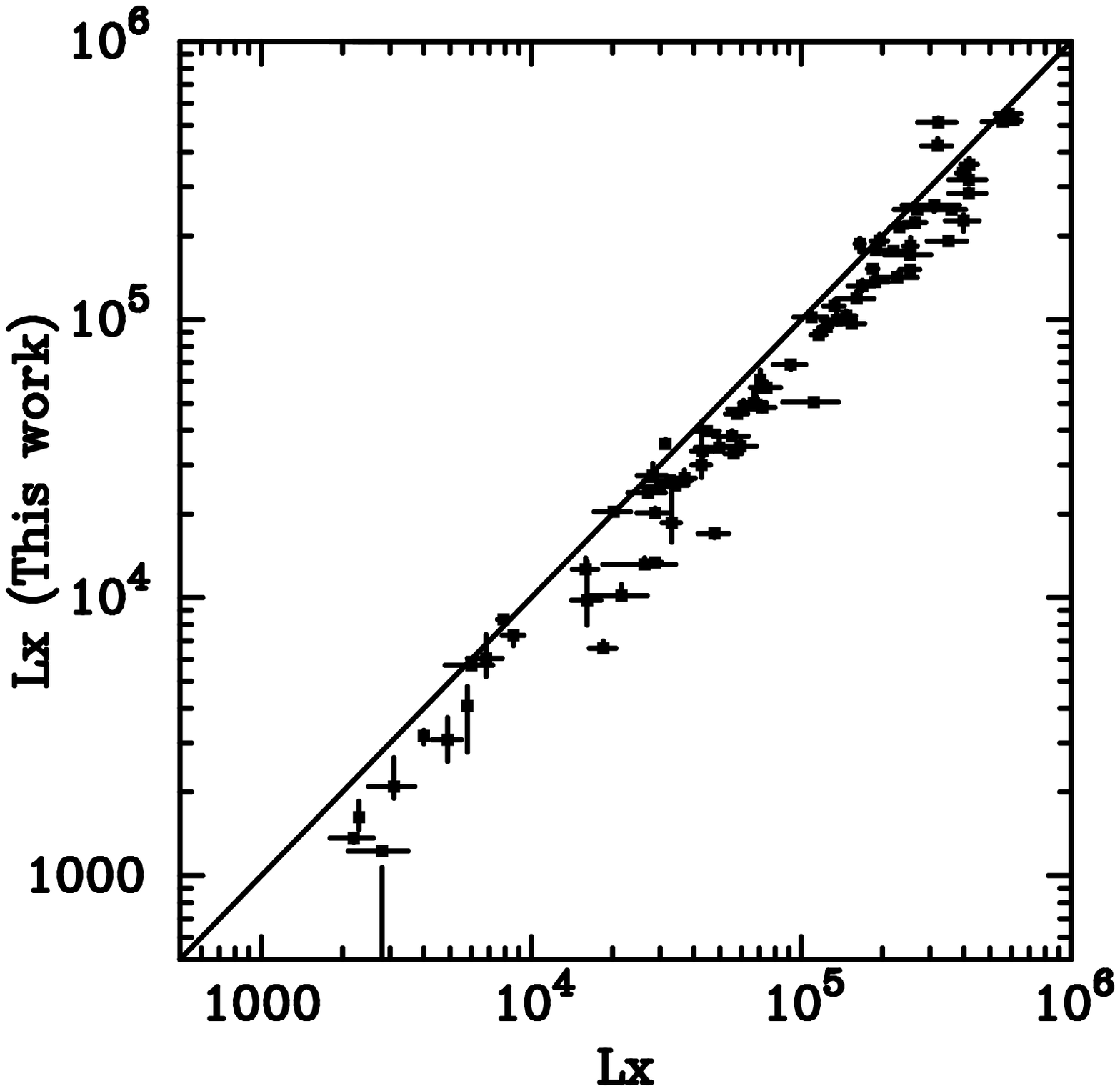}}
\caption{Comparison of the bolometric X-ray luminosity of our results
 with those in Xue and Wu (2000).}
\label{lxcmp}
\end{minipage}\quad
\begin{minipage}[tbhn]{8cm}
\centerline{\includegraphics[width=9.5cm]{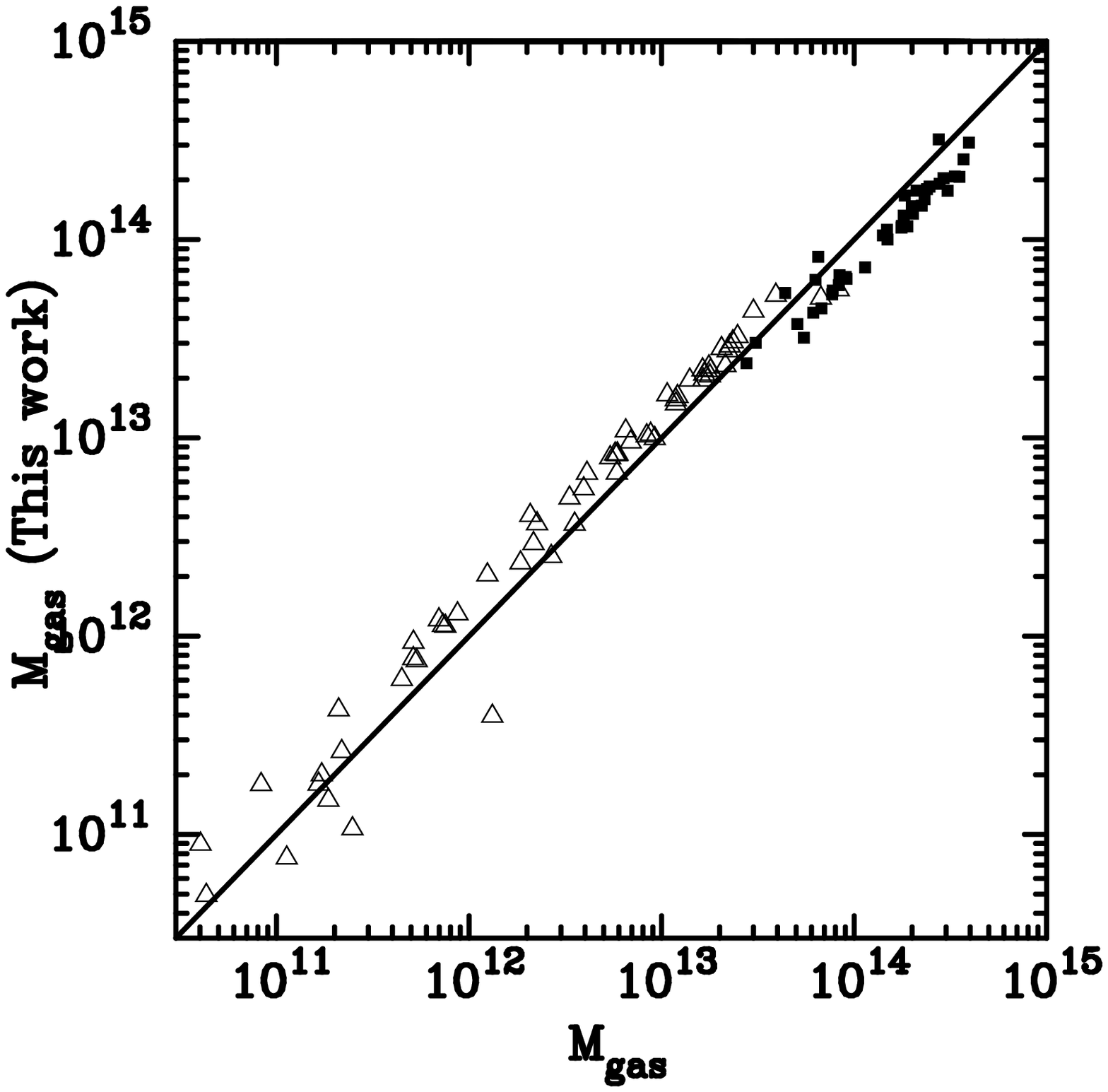}}
\caption{Comparison of the gas mass of our results with those in Mohr,
 Mathiesen, and Evrard (1999) and Sanderson et al. (2003), 
 for the solid square and open
 triangle, respectively. The integration radius was chosen to be
 the value in these two references. }
\label{gmascmp}
\end{minipage}
\end{figure}

We derived an observed X-ray flux, $F_{\rm X}$, and luminosity, $L_{\rm
X}=4\pi D^2F_{\rm X}$, 
in the 0.5--10 keV energy range in the rest frame, 
by integrating the simulated emission 
spectra of the best-fit model within the best-fit $R_{\rm limit}$.
In order to consider the emission out of the GIS field of view 
for objects with $R_{\rm limit}>18'$, we integrated the flux up to the
$R_{\rm limit}^{\rm av}(kT)$, which is an average detection radius for
the specific temperature (figure \ref{t-rdet}): $0.2(kT/{\rm
1keV})^{1.70}$ Mpc for $kT<3$ keV and $1.3(kT/{\rm
3keV})^{0.823}$ Mpc for $kT>3$ keV.
This correction is typically at most 1.0--1.2, but $\sim$2.0 for several
nearby bright clusters, such as Centaurus, Perseus, and Virgo.
We also calculated the bolometric luminosity by using the emissivity of
the plasma model in the 0.001--200 keV band in the rest frame.
For the double-$\beta$ model, we derived it separately for the small 
and large-scale components.
In order to estimate the error range of each quantity,
we fitted the radial count rate profile by the $\beta$ model
for all of the imaging parameter sets within the 90\% confidence range.
We then calculated the flux and luminosity for each parameter set, and
obtained the permitted range of each quantity.
The obtained results of flux and luminosity, together with the central
X-ray surface brightness, are given in table \ref{fxlx-lst}.
The errors are typically less than 15\%, which are primarily 
attributed to the uncertainties of $\beta$, the core radii ($a_1$, $a_2$), and $R_{\rm limit}$.
In figure \ref{lxcmp}, 
we compare the bolometric luminosity obtained here with that in Xue and
Wu (2000), who collected the available results of various authors.
Both correlate well, but our results are systematically lower by $\sim$20\%.
One reason is that Xue and Wu (2000) corrected the observed luminosity
by taking into account the undetected emission at the periphery, while
our results were calculated within the detection-limited radius.
This effect explains the difference of 10--15\%.

We checked the consistency for the ICM mass between ours and 
the previous results of Mohr, Mathiesen, and Evrard (1999) and 
Sanderson et al. (2003).
The integration radius was chosen from this literature for the same
objects analyzed here.
Since the ICM mass, itself, is not available in Sanderson et al. (2003), we
calculated it from the $\beta$-model parameter in their paper.
The ICM mass density was calculated from $\rho_{\rm gas}(r)=\mu_{\rm e}
m_{\rm p} n_{\rm e}(r)$,
where we assumed $\mu_{\rm e}=1.167$, 
based on the single-$\beta$ model fitting.
In figure \ref{gmascmp}, we compare our results with those of two references.
The systematic difference is seen against both references, although our
results are in the middle of the difference between the two papers.

The derivation of the ICM mass is not simple, because it strongly depends 
on the integration radius.
The well-defined radius (Evrard 1997) $r_{500}$, within which the average 
mass density is 500-times as high as the 
critical density of $\frac{3H_{0}^2}{8\pi G}=1.56\times10^{-29}h_{50}^2$ g cm$^{-3}$,
has often been used to integrate the ICM mass.
However, the X-ray emission is not always confirmed up to $r_{500}$, 
furthermore, no
information has been obtained about the ICM temperature up to $r_{500}$.
These results do not ensure us to derive the ICM mass.
Therefore, in addition, 
we introduced an alternative radius, $r_{1500}$, within which
the averaged
mass density is 1500-times as high as the critical density.
As shown in 4.4, this radius is as large as $0.3r_{200}$, 
used in Sanderson et al. (2003).
We obtained the relation ${\bar r_{1500}}=0.291\left(\frac{kT}{\rm 1
keV}\right)^{0.657}h_{50}^{-1}$ Mpc from our sample clusters for $kT>1$ keV, 
and then calculated $M_{\rm gas}({\bar r_{1500}})$ as 
\[M_{\rm gas}({\bar r_{1500}})=\int 4\pi r^2 \rho_0 dr=\int 4\pi r^2
\mu_{\rm e} m_{\rm p} \sqrt{\sum_{i=1}^{k}n_i^2}dr\qquad\qquad\mbox(k=1,2)\]
for each of our sample clusters.

\section{Correlation Studies}

\subsection{LT Relation}

\begin{figure}[hptb]
\begin{minipage}[tbhn]{8cm}
\centerline{\includegraphics[width=9.5cm]{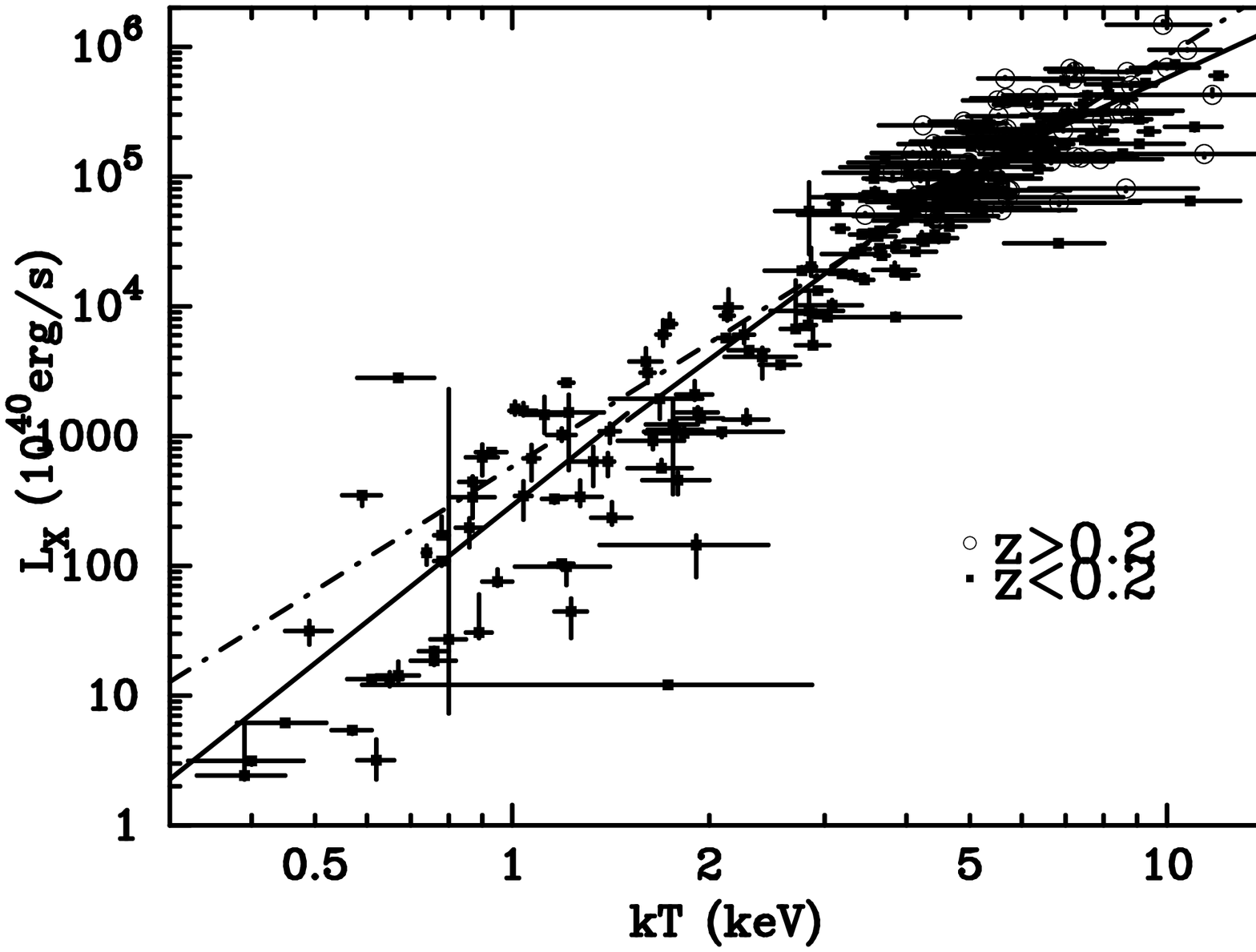}}
\end{minipage}\quad
\begin{minipage}[tbhn]{8cm}
\centerline{\includegraphics[width=9.5cm]{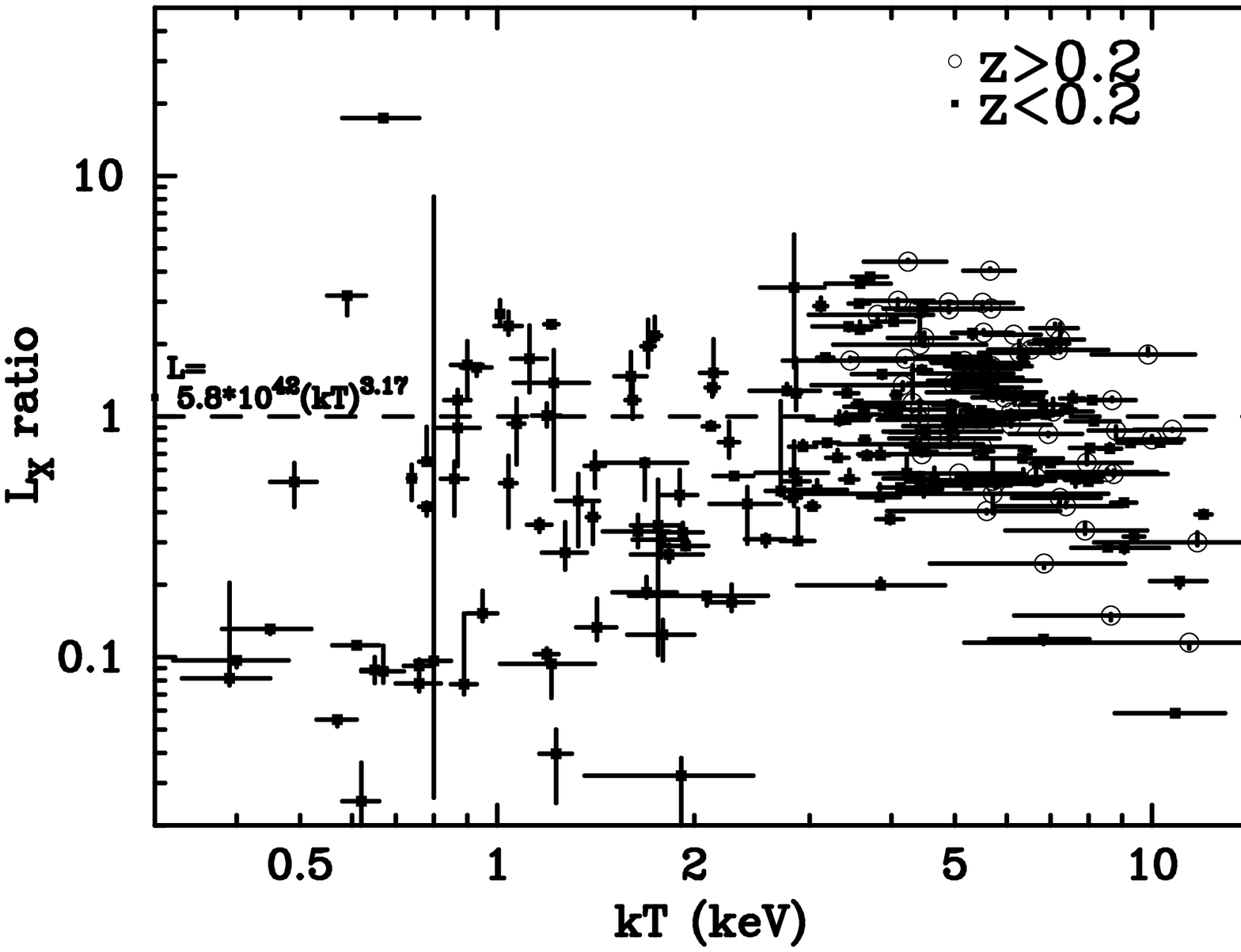}}
\end{minipage}
\caption{Left panel is a plot of the
 X-ray bolometric luminosity against the gas
 temperature. The three solid lines represent the best-fit
 relations in the energy range of 0.3--1.5, 1.5--5, and 4--15 keV, and
 the dashed line is that in 0.3--15 keV. See the text for the
 equations. The right panel is the luminosity ratio against the best-fit
 single power-law relation ($L=5.8\times10^{42}T^{3.17}$).}
\label{tlx}
\end{figure}

First, we consider the temperature and X-ray luminosity relation,
which is a basic relation on clusters.
In figure \ref{tlx} (left), we plot the bolometric luminosity against the gas
temperature.
Since bright distant clusters were preferably proposed to be observed,
luminous clusters are dominant in number.
Nevertheless, there are enough samples of low-temperature objects
continuously down to $kT=$0.4 keV.
We can find no clear difference in the LT relation 
between nearby (redshift $z$ is $<0.2$) 
and distant ($z>0.2$) clusters, as claimed by Mushotzky and Scharf
(1997), who analyzed the ASCA data.
In the following, we consider the LT relation by excluding objects whose
accuracy of temperature is $>30$\% and redshift is $>0.5$.
When there are objects whose luminosity is by a factor of $<0.2$ or $>5$
higher than the obtained relation, 
we exclude them and again obtain the relation.

First, in order to obtain the overall LT relation of galaxy clusters and
comapre it with the past results, we fit the data at temperatures of
1.5--15 keV, by the WLS method (Akritas, Bershady 1996).
As a result, we obtained the relation
$L_{\rm X}=5.80_{-1.26}^{+1.61}\times10^{42}(kT)^{3.17\pm0.15}$ erg s$^{-1}$, where the temperature, $kT$, is in units of keV.
This is somewhat steeper than the index of $2.72\pm0.05$ reported by Wu,
Xue, and Fang (1999), 
and near to $3.03\pm0.06$ in Xue and Wu (2000), who include both rich
clusters and galaxy groups.
When we chose only objects, whose luminosity is in a similar range of 
Wu, Xue, and Fang (1999), $>10^{44}$ erg s$^{-1}$, the index becomes
$2.61\pm0.20$, close to the value of Wu, Xue, and Fang (1999).
In our plots, the index below 1.5 keV becomes $4.03\pm1.07$, 
consistent with $5.30\pm1.79$ in Xue and Wu (2000) and $4.9\pm0.8$ in
Helsdon and Ponman (2000).
Therefore, the overall relation of our results follows the previous
results, in terms of the steepness of the relation and one break point 
around $kT\sim1$ keV.
Note that
X-ray faint objects around $10^{40-41.5}$ erg s$^{-1}$ and $kT\sim$0.4--0.8
keV are mainly
elliptical galaxies in our sample, while they are galaxy groups in the 
previous studies.
This indicates that elliptical galaxies and galaxy
groups have the same LT relation.

There are, however, several clusters whose luminosities are significantly
below the typical relation around $kT\sim$2--3 keV (NGC 4756, NGC 3258, 
RXJ 1833.6+652), 6--7 keV (A 2556B), and 10--13 keV (A 2556A); their
luminosities are 5--10 times lower than the typical relation.
The X-ray emission of A 2556 is doubly peaked, and the merging is now progressing,
indicating that the temperature becomes higher by shock.
The reason for the low luminosity of poor clusters NGC 4756 and NGC 3258 is
not obvious; they might belong to the class objects on the 
extraporation of X-ray faint galaxy groups.
The results of RXJ 1833.6+652 might be contaminated by the background
distant cluster associated 3C 383 (z=0.161; NED).
On the contrary, one cluster, A 1885 ($kT=$2.3 keV), is especially highly
luminous for its temperature by a factor of 5--10.

On the other hand, when looking at the plot in detail, we can see a hint
of another break point of the LT relation around 4 keV.
Such a break can be seen more clearly in figure \ref{tlx} (right), where
we plot the luminosity ratio against the best-fit single power-law relation. 
When we obtained the steepness index separately above and below 4.5 keV,
it became 2.34$\pm$0.29 for $4<kT<15$ keV and 3.74$\pm$0.32 
for $1.5<kT<5$ keV.
Considering this large difference of steepness, and especially a large
scatter of luminosity around $kT\sim$3--4 keV, it is suggested that two
distinct classes of objects exist; one belongs to
the same class as richer clusters, and the other to poorer clusters; two
classes coexist around $kT=$3--4 keV where higher luminosity clusters
are at the cooler end of rich clusters and lower ones are at the hotter
end of poor clusters.
Such a break-like feature can be seen in the LT relation of Xue and Wu 
(2000).

\subsection{X-ray Distribution and Temperature}

The temperature dependence of the X-ray distribution of hot gas is an important
property to consider the evolution of clusters and galaxies, since
lower temperature systems exhibit a variety of characteristics, and
it seems that they cannot be explained by a simple picture.
We could derive a key issue when we understand how 
they connect to rich clusters concerning the X-ray properties.
It has been reported that the lower temperature systems exhibit a
smaller value of $\beta$ and core radius in the $\beta$-model fitting
(Mohr, Mathiesen, and Evrard 1999; Sanderson et al. 2003);
this leads to the claim of an entropy floor (Ponman et al. 1999).
Matsushita (1997, 2001), Beuing et al. (1999), Brown and Bregman
(2000) indicated two different classes of
elliptical galaxies in terms of the X-ray size.
The ICM-to-stellar mass ratio is reported to be smaller for these
systems than rich clusters.
In order to understand these properties in a unified picture, we
investigated the temperature dependence of the X-ray distribution from
elliptical galaxies to rich clusters simultaneously.

In figure \ref{betafit} (left), 
we plot the results of a single $\beta$-model fitting against
the gas temperature.
It can be seen that the $\beta$ and core radius strongly correlate 
with the temperature.
This is in good agreement with the result of Horner, Mushotzky, and
Scharf (1999) and 
Sanderson et al. (2003).
On the contrary, the correlation between the central electron density and
temperature is not clear.
Probably results of single-$\beta$ model fittings could be affected by
the double structure of the X-ray surface brightness profile which is
frequently seen for galaxy groups (Ikebe et al. 1996; Mulchaey, 
Zabludoff 1998); when the surface brightness of the double-$\beta$
structure is fitted with the single-$\beta$ model, the parameters
$\beta$ and core radius becomes smaller.
In fact, as show in figure \ref{imfit}, some poor clusters apparently exhibit a
double-$\beta$ model structure.
The double-$\beta$ model is physically reasonable, since the depth of 
the gravitational potential of the central galaxy is comparable to that of
poor clusters, and thus the
hot gas bound by the central galaxy becomes relatively dominant at the
cluster center.
In figure \ref{betafit} (right), 
we show plots in which we replace the results of a single-$\beta$ 
model fitting by those of outer component in the double-$\beta$ 
model fitting for the clusters listed in table \ref{b2fit-lst}.
Compared with figure \ref{betafit} (left), 
the correlation between $\beta$ and the core radius with
the temperaure becomes somewhat small, but still remains.
This trend is also reported by Helsdon and Ponman (2000), who used
the double-$\beta$ model to fit the surface brightness.
The most remarkable difference is for the central electron density;  the
correlation with the temperature becomes clearer in figure \ref{betafit}
(right), in such a way that the correlation coefficient is 0.077 and 0.431
for plots of the single and double-$\beta$ model fits, respectively.
In other words, the lower is the temperature, the lower is the central
electron density.
At the lowest temperature, some objects do not show the double-$\beta$
structure. 
The $\beta$ value is almost the same as that of groups and clusters, while the
core radius is by an order of magnitude smaller.
The central electron density is significantly higher than that of
galaxy groups.
These objects are X-ray faint elliptical galaxies.
The difference of the X-ray surface profile between galaxies and clusters is
also indicated by Sanderson et al. (2003).
These results indicate that we see two distinct types of hot gas, galaxy and
cluster components.
The cluster component is dominant in rich clusters, and not seen in
X-ray faint elliptical galaxies.
The galaxy component is seen in elliptical galaxies and poor clusters,
and not seen in rich clusters.
In addition, a weak correlation between the core radii and the temperature 
shown in figure \ref{betafit} (left)
may be possibly due to the hot gas in the central elliptical galaxy.

In summary, the X-ray distribution of hot gas depends on the
temperature; the $\beta$ and core radius are smaller for lower
temperature clusters, even when considering the double-$\beta$ model structure.
This trend agrees with the previous studies, based on the ROSAT data.
On the other hand, in considering the central electron density, 
we must take into account the gas component, which is bound by the 
central galaxy.
When looking at only the cluster component, the central electron density is
smaller for lower temperature clusters.

\begin{figure}[hptb]
\vspace*{1cm}
Figures are shown at the end of this text.
\caption{Gas-temperature dependence of the $\beta$-model
 parameters. The top, middle, and bottom panels show $\beta$, the core
 radius, and the central electron density, respectively. The left panels 
 plot the results of single-$\beta$ model fittings. The right panels
 plot the results of double-$\beta$ model fittings for the objects listed in
 table \ref{b2fit-lst} (open circle), 
 together with those of single-$\beta$ model
 fittings for other objects (filled square).}
\label{betafit}
\end{figure}

\subsection{The Maximum Detection Radius of X-ray Emission}

So far, the X-ray size and temperature (ST) relation has been reported by
Mohr and Evrard (1997); 
the maximum X-ray size is proportional to the temperature.
This study was mainly performed for higher temperature clusters.
On the other hand, Matsushita (1997, 2001) claimed a large difference of
X-ray size between X-ray bright and faint elliptical galaxies.
In addition, Ponman et al. (1996), Helsdon and Ponman (2000), and
Mulchaey (2000) insisted on the necessity of
considering the undetected X-ray emission for X-ray faint galaxy groups.
We thus have great interest in the connection between these 
classes of objects.
The maximum radius is, of course, dependent on the sensitivity of
the detectors, but
our systematic study on the X-ray size of hot gas with the same instrument
therefore has a great advantage of avoiding this effect.
Strictly speaking, the estimation of the maximum radius is also dependent 
on the exposure time.
The exposure time is typically in the range of 15--60 ks, and such
a difference introduces an error difference by a factor of $\sim$2, and
thus the difference of the maximum radius by a factor of $\sim1.4$ for the 
$\beta$-model profile with $\beta=0.5$, when ignoring the background
error.
Therefore, considering the background error,
the difference is smaller than 1.4.
As can be seen from the later, 
such a difference is not significant in our studies.
What we have to pay attention to is an observed energy band, because of a
wide range of object temperature.

\begin{figure}[hptb]
\begin{minipage}[tbhn]{8cm}
\centerline{\includegraphics[width=9.5cm]{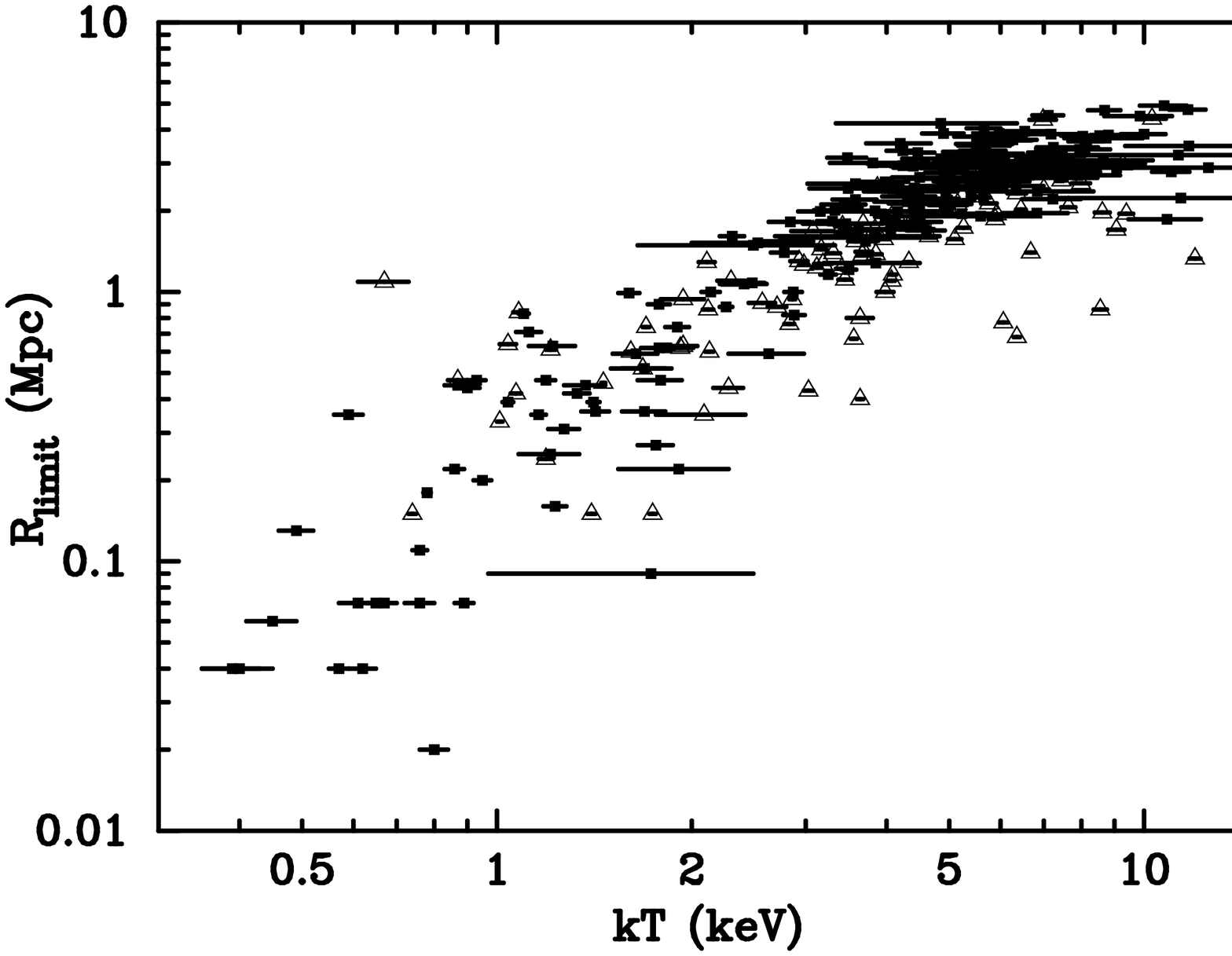}}
\caption{Maximum detection radius, $R_{\rm limit}$, 
against the gas temperature. The triangles represent the lower limit for
 objects whose emission is extended beyond the GIS field of view.}
\label{t-rdet}
\end{minipage}\quad
\begin{minipage}[tbhn]{8cm}
\centerline{\includegraphics[width=9.5cm]{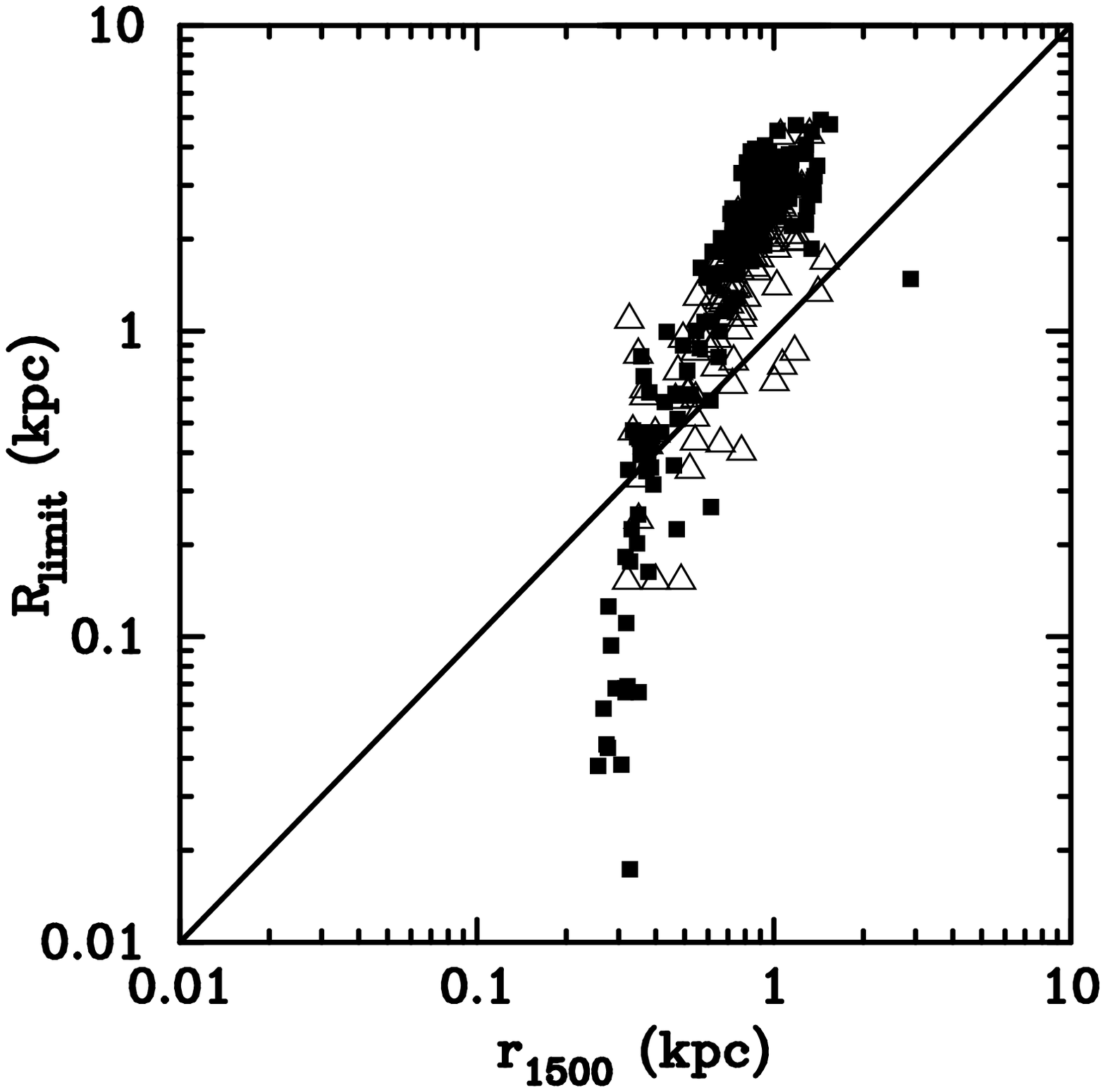}}
\caption{Comparison of $R_{\rm limit}$ and $r_{1500}$. 
The triangles represent the lower limit of $R_{\rm limit}$ 
for objects whose emission is extended beyond the GIS field of view.}
\label{rcmp}
\end{minipage}
\end{figure}

The maximum radius of X-ray size corresponds to $R_{\rm limit}$, as
treated in subsection 3.5.
In figure \ref{t-rdet}, we plot them against the temperature.
Since the X-ray emission of nearby objects is often larger than the GIS
field of view, we denote such objects by circle symbols as a lower limit.
For rich clusters, the size is as large as 3--5$h_{50}^{-1}$ Mpc, while
low-temperature objects have a size of only 50--300$h_{50}^{-1}$ Mpc.
On the whole, the X-ray size, $R_{\rm limit}$, correlates with the
temperature, as reported by Mohr and Evrard (1997).
In addition, we can clearly see a novel feature: 
two breaks at around $kT=$1 and $kT\sim$3--4 keV.
Such breaks are similar to those seen in the LT relation, 
and $R_{\rm limit}$ scatters largely below 2 keV, in contrast with 
high-temperature objects.
Therefore, the LT relation is thought to greatly concern the maximum
X-ray size.

The size of galaxy clusters is often defined as the radius,
$r_{500}$, although the X-ray emission has not been confirmed up to this
radius for low-temperature objects.
Then, we compare the X-ray maximum radius $R_{\rm limit}$ 
with $r_{1500}$ which is introduced in subsection 3.7.
As can be seen in figure \ref{rcmp}, the 
relative size of $R_{\rm limit}$ against
$r_{1500}$ is quite different for various temperature.
For high-temperature objects whose $r_{1500}$ is larger, $R_{\rm limit}$
is $\sim3$ times larger than $r_{1500}$.
On the other hand, $R_{\rm limit}$ is comparable to, or much smaller 
than, $r_{1500}$ for lower-temperature objects.
These trends are in good agreement with the comparison between 
$r_{500}$ and ROSAT $R_{\rm limit}$ (Mulchaey 2000).
The small $R_{\rm limit}$ of low-temperature objects is not due to the 
low detection efficiency of the GIS for the soft
X-ray emission; the GIS count rate for the same emission measure 
is smaller by a factor of at
most 2 for $kT=$0.5 keV, compared to $kT=$1 keV, and it corresponds to
a reduction by 1.4 for the detection radius for the count rate 
radial profile of $r^{-2}$, assuming $\beta=0.5$.
Since $r_{1500}$ correlates well with the temperature (subsection 3.7), this
trend can be indicated from figure \ref{t-rdet}.
This is strong evidence that the X-ray properties of galaxy clusters
do not simply follow the scaling law, and the X-ray size of 
low-temperature objects is much smaller than the theoretical prediction.
We note that X-ray faint elliptical galaxies smoothly connect with
galaxy clusters without a significant transition, indicating the
possibility that the X-ray properties of X-ray faint ellipitcal 
galaxies and rich
clusters can be explained within the same framework by a unified picture.

\subsection{The Gas Mass and Temperature Relation}

In treating the gas mass, we must define the integration radius, 
which should depend on the system temperature.
Since we have no obvious definition of the integration radius, 
as described in the previous section, 
we investigated the gas mass within the two radii, $R_{\rm
limit}$ and $r_{1500}$.
For $r_{1500}$, we did not apply a measured value of each object, but
a calculated value from the observed relation of 
$r_{1500}=0.358\left(kT/{\rm 1 keV}\right)^{0.554}$ Mpc, so as to avoid
any error concerned with the observed $r_{1500}$.
This integration radius is as large as that used by Sanderson et
al. (2003), $0.3r_{200}=0.29\left(kT/{\rm 1 keV}\right)^{0.5}$ Mpc.

\begin{figure}[hptb]
\begin{minipage}[tbhn]{8cm}
\centerline{\includegraphics[width=9.5cm]{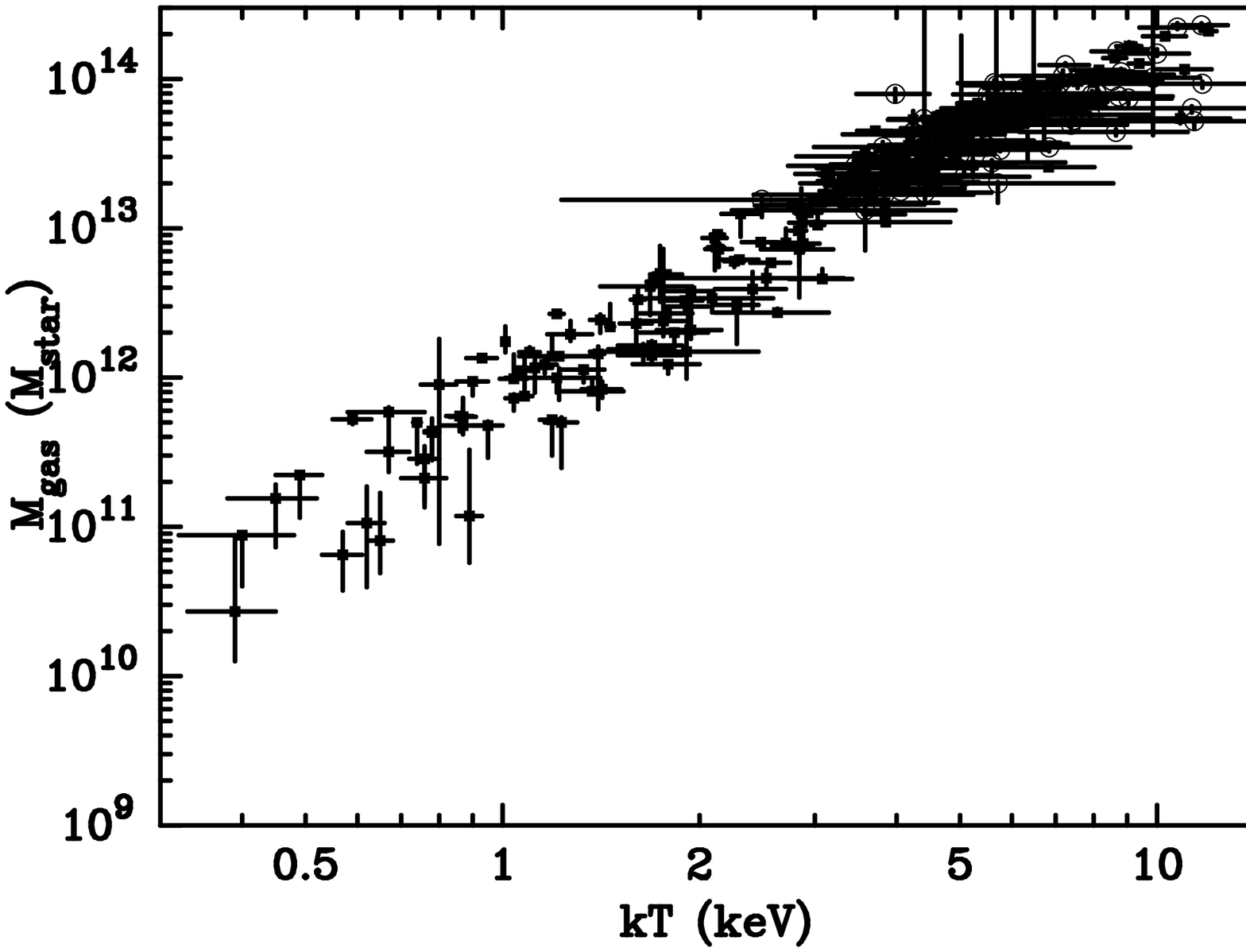}}
\end{minipage}\quad
\begin{minipage}[tbhn]{8cm}
\centerline{\includegraphics[width=9.5cm]{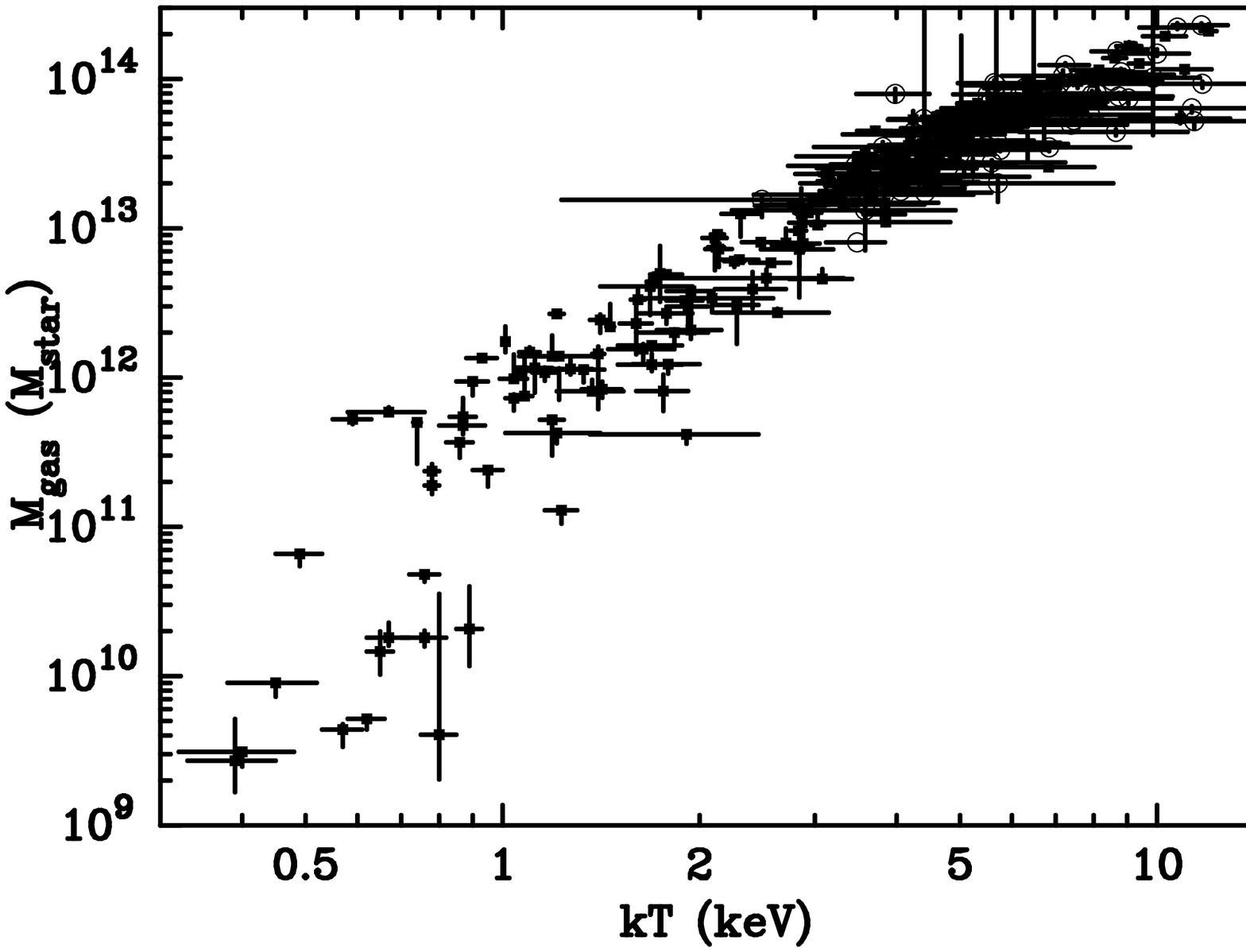}}
\end{minipage}
\caption{Gas mass against the gas temperature. Left plots the
 mass within $r_{1500}$, and right plots one within $R_{\rm limit}$
 for objects whose $R_{\rm limit}$ is smaller than $r_{1500}$, together
 with the mass within $r_{1500}$ for other objects. 
 The symbols are the same as those in figure 12. }
\label{gmass}
\end{figure}

First, figure \ref{gmass} (left) is a plot of the gas mass, $M_{\rm
gas}(r_{1500})$, against the temperature.
The $M_{\rm gas}(r_{1500})$ correlates well with the temperature, and
smoothly connects as a line from rich clusters to galaxy groups.
X-ray faint elliptical galaxies seem to locate slightly below this 
relation by a factor of 2--3, but it is ambiguous.
Above $kT>1$ keV, the relation is represented by $M_{\rm
gas}(r_{1500})=(1.0\pm0.1)\times10^{12}\left(kT/{\rm 1 keV}\right)^{2.33\pm0.07}$.
This indicates that one parameter, such as the temperature, determines the
gas mass from rich clusters to elliptical galaxies. 
The slope of 2.33 is almost similar to the value 1.98$\pm$0.18 for 
the gas mass within $r_{500}$ (Mohr et al. 1999).

However, we should not accept the above relation straightforwardly, since we
cannot confirm that the X-ray hot gas is extended up to $r_{1500}$ in groups
and galaxies, as shown in the previous section.
When we replace $M_{\rm gas}(r_{1500})$ by $M_{\rm gas}(R_{\rm limit})$
for objects whose $R_{\rm limit}$ is smaller than $r_{1500}$, the
relation becomes as figure \ref{gmass} (right).
The gas mass of low-temperature objects dramatically decreases, and thus
the relation has a break at around 1 keV.
Another break at around 3--4 keV is seen as the LT relation, although
this also appears in figure \ref{gmass} (left).
Considering figure \ref{gmass}, it is claimed that 
the gas content within the viriral radius is well scaled by the system 
temperature, while there is a difference 
in the gas distribution between lower and higher temperature
objects.

\begin{figure}[hptb]
\centerline{\includegraphics[width=9.5cm]{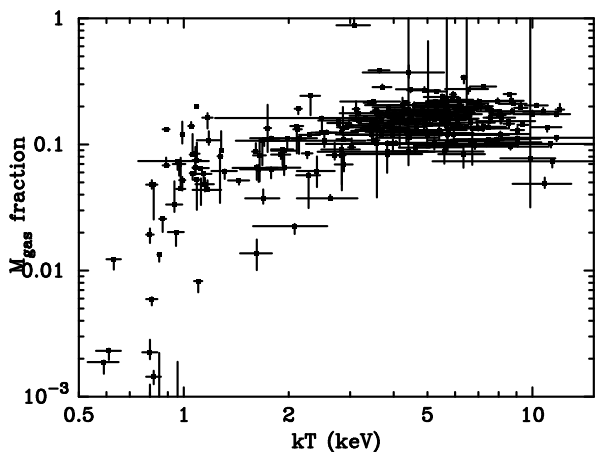}}
\caption{Gas mass fraction against the gas temperature within
 $r_{1500}$. For objects whose $R_{\rm limit}$ is less than
 $r_{1500}$, the fraction within $R_{\rm limit}$ is plotted.}
\label{gmassfrac}
\end{figure}

We are also interested in the scatter of the gas mass fraction.
In figure \ref{gmassfrac}, we plot the gas mass fraction against the 
temperature.
For a higher temperature of $kT>4$ keV, the fraction narrowly distributes 
in the range of 0.1--0.2$h_{50}^{-1.5}$.
As claimed by Loewenstein and Mushotzky (1996), a significant
scatter of the gas mass fraction within a specific radius 
indicates that the evolution of galaxy clusters is governed by not only
gravity, but also other mechanisms.
The fraction slightly, but significantly, decreases toward the lower
temperature, and it has a break at around $kT\sim1$ keV.
This sharp break is enhanced evidence of non-gravity effects 
on the evolution.

\begin{figure}[hptb]
\begin{minipage}[tbhn]{8cm}
\centerline{\includegraphics[width=9.5cm]{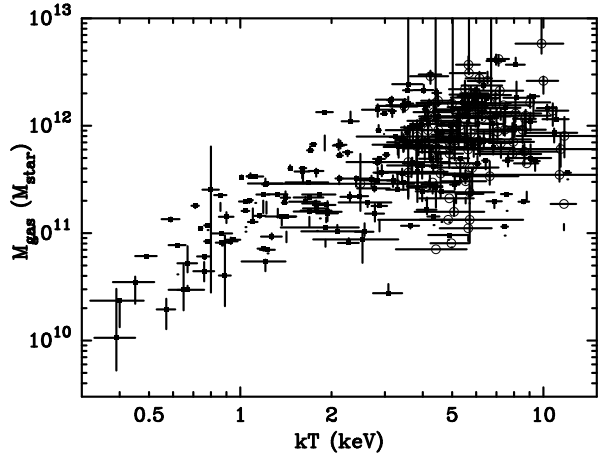}}
\end{minipage}\quad
\begin{minipage}[tbhn]{8cm}
\centerline{\includegraphics[width=9.5cm]{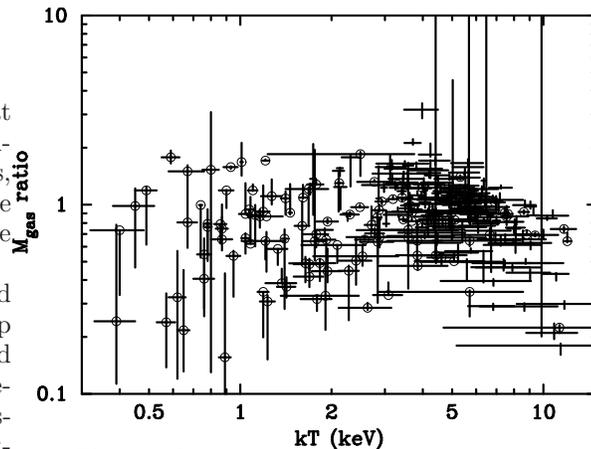}}
\end{minipage}
\caption{Gas mass against the gas temperature. The left panel plots the
 gas mass within 120$h_{50}^{-1}$ kpc. The symbols are the same as
 those in figure 12. 
 The right panel plots the ratio of the gas mass of individual objects 
 within $r_{1500}$ to
 the cluster-averaged one calculated from the relation of
 $1.0\times10^{12}\left(kT/{\rm 1 keV}\right)^{2.33}$, 
 by denoting objects
 whose gas mass within 120$h_{50}^{-1}$ kpc is less than $5\times10^{11}
 M_{\odot}$ as open circles.}
\label{gmass120}
\end{figure}

Last, let us look into the relation between the central gas density and
the overall gas mass.
The central gas density is an indicator of ``cooling flow rate'',
and is thought to be related to the cluster evolution history.
Here, we took the gas mass within 120$h_{50}^{-1}$ kpc of the cluster
center instead of the central gas density, because of large uncertainties for
the central electron densities.
This is why we chose a radius of 120$h_{50}^{-1}$ kpc; this value is
a median of the core radius in rich cluster with $kT>4$ keV [figure
\ref{betafit} (left middle)].
In figure \ref{gmass120} (left), the gas mass within 120$h_{50}^{-1}$ kpc
is plotted.
The gas mass is not constant, but widely distributed by an order of
magnitude.
There is a systematic difference between lower and higher temperatures; 
this feature should be related to that in the LT relation.
We classify our sample by this central gas mass, below and above 
$5\times10^{11} M_{\odot}$, and show the ratio of the gas mass 
within $r_{1500}$ to the
cluster-averaged one calculated from the relation of
$1.0\times10^{12}\left(kT/{\rm 1 keV}\right)^{2.33}$", in figure \ref{gmass120} (right).
Most of low-temperature objects belong to the low-central-density class.
Interestingly, objects with a lower central gas mass exhibit a lower
gas mass within the virial radius at $kT\sim$3--5 keV, 
indicating that the gas density of
a lower-central-mass object is lower at all radii than a
higher-central-mass object.
On the other hand, no difference is seen above $kT>5$ keV between both
types of objects.
In analogy with the case of the LT relation, two classes of hot gas
systems appear; 
the low-central-mass
object corresponds to the class below 3--4 keV with a steep LT
relation, while the high-central-mass one cprresponds to rich clusters
with a flatter LT relation.

\section{Discussion}

We analyzed about 300 objects of galaxy clusters, galaxy groups, and
elliptical galaxies systematically with the ASCA data, especially the GIS data.
Our results are free from systematic uncertainties, such as 
differences between instruments and analysis procedures.
In addition, our sample includes a wide range of system temperatures
from 0.4 keV to $\sim12$ keV.
Utilizing this sample, we performed correlation studies, and found
several interesting phenomena.
Here, we discuss them from the view point of evolution, hierarchical
structure, and so on.

\subsection{Two Breaks in the LT Relation}

We found that the LT relation exhibits two breaks at around 1 keV and 3--4
keV.
These two breaks are also indicated by the temperature dependence of
other properties, such as the hot-gas mass, size, and gas fraction.
Here, we discuss the origin and implication of the two breaks in the
temperature dependences.

The break at around $kT\sim1$ keV was already reported 
by Ponman et al. (1996), and so
on, as well as the steep LT relation below $kT=1$ keV.
Ponman et al. (1996) and Helsdon and Ponman (2000) 
suggested that this steep relation is
due to galaxy feedback; lower-temperature systems were more strongly
affected by non-gravity heating, and the gas has escaped or become thinner.
Balogh, Babul, and Patton (1999) reproduced this steep relation in 
semi-analytical models.
Alternatively, as described in the previous subsection, it is suggested that
the X-ray emission of low-temperature objects is around the detection
threshold, and is possibly lost at the outermost region for 
low-surface-brightness objects, and thus the X-ray luminosity may be underestimated
(Mulchaey et al. 2000).
Accordingly, the slope of the LT relation in lower temperature 
objects may possibly become as small as that of galaxy clusters.
In fact, if we estimate the X-ray luminosity within the radius
$r_{1500}$, rather than the detection radius, $R_{\rm limit}$, 
the luminosity of low-temperature objects significantly increases, 
while the luminosity of high-temperature objects slightly decreases.
This behavior can be understood from figure \ref{rcmp}. 
The slope of the LT relation becomes $2.56\pm0.33$, $3.68\pm0.37$, and 
$3.40\pm0.59$, 
for ICM temperatures of $>4$ keV, 1.5--5 keV, and $<1.5$ keV,
respectively. 
Thus, the break at $kT\sim$1 keV disappears.
If this is true, 
the existence of undetected group-scale X-ray emission is necessary 
around X-ray faint elliptical galaxies; we discuss this possibility
in the next subsection.
We infer that the hot gas escaped from low-temperature
objects and some fractions remain, but cannot be easily detected.

The second break at around $kT\sim$3--4 keV is inferred from the other ICM
features.
The entropy and temperature relation shown by Ponman, Cannon, and
Navarro (1999)
exhibits an excess entropy below 3--4 keV.
The Si-to-Fe abundance ratio of the ICM is constant above 4 keV, while
it decreases toward the lower temperature below 4 keV (Fukazawa et
al. 1998).
The iron-to-stellar mass ratio also follows the same trend (Fukazawa
1997; Finoguenov et al. 2000, 2001), 
reflecting the trend of the gas mass fraction that
we present in this paper (figure \ref{gmassfrac}).
These three issues can be explained by the scenario that a significant
amount of ICM was pushed out by the galactic
wind at the early galaxy formation epoch, driven by the burst of 
a type-II supernovae.
Accordingly, it is reasonable that the break point around $kT=$3--4 keV
is seen in the LT relation.
The slope of the LT relation above $kT>4$ keV is
2.34$\pm$0.29, close to 2.0 predicted by the theoretical scaling law, 
while that for objects with $1.5<kT<4$ keV is 3.74$\pm$0.10, 
significantly steep.
Furthermore, a large scatter around $kT\sim$3--4 keV implies the
coexistence of objects belonging to the higher-temperature classes 
in the flat LT relation and to the lower temperature in the steep LT
relation.
Considering these features, it is suggested that higher temperature
objects are well-relaxed systems, and their non-gravity effect becomes
negligible due to the large amount of gravitational energy.
As a result, they follow a scaling law in the LT relation.
On the other hand, lower-temperature objects have significantly
sufferred non-gravity heating, such as galactic winds, and thus do
not follow the scaling law.
This scenario also explains the fact that there is 
a correlation of the gas mass between the center and outer regions
(figure \ref{gmass}).
The non-gravity effect prevented the gas from condensing in the
potential, and, as a result, the gas density is lower over the
cluster-scale.

In the bottom-up scenario, rich clusters with higher temperature are formed
through mergers of poorer clusters.
Such poorer objects are thought to belong to the lower-temperature
classes with excess entropy.
After merging, the excess entropy may be removed in the release of
extra-gravitational energy by radiation and so on to achieve virial
equilibrium.
Cavaliere, Menci, and Tozzi (1997, 1999) presented the simulation and analytic
calculation of the cluster evolution, where they considered the
preheating, subsequent hierarchical merging, shock heating/compression,
and a new hydrostatic equilibrium.
They found that the slope of the LT relation is 5 for galaxy groups, 3 for
rich clusters, and saturates toward 2 for the highest-temperature clusters.
This trend is in good agreement with our results, and indicates that
smaller systems suffer the merger effect, while larger systems do not, 
because of few merger events where comparable subclusters mix with eath other.

It has been said that the LT relation of clusters of galaxies is steeper
than that predicted by the scaling theory, and thus several alternative
scenarios are proposed, which take into account gravitational heating,
radiative cooling, and so on (Muanwong et al. 2002; Tornatore et al. 2003).
Most of these scenarios predict a steep LT relation over any cluster
temperature.
However, our results show that rich clusters with $kT>4$ keV at least
satisfy the scaling relation.
As shown in subsection 4.2, 
the hot-gas distribution of cluster-scale components 
is less concentrated for lower-temperature objects.
Heating or cooling is a possible effect to explain this trend.
Since not all of the rich clusters exhibit evidence of radiative cooling,
we think that the heating effect is preferable.
In order to reproduce our results, heating effects should not be so
large as to have influence on the LT relation above $kT>4$ keV.

\subsection{Gas Distribution and Two Component Models}

Concerning the X-ray surface brightness, we found that many objects exhibit a
double-$\beta$ model structure, especially for lower-temperature
objects, while rich clusters and X-ray faint elliptical galaxies do not.
Such a double-$\beta$ structure has already been found by Ikebe et
el. (1996) and Mulchaey and Zabludoff (1998) for galaxy groups, and by
Matsushita et al. (1998) for the X-ray bright elliptical galaxy NGC 4636.
Even for rich clusters, Ikebe et al. (1997) and Xu et al. (1998)
have suggested a double-$\beta$ structure, although the poor angular resolution
of ASCA cannot distinguish between a double-$\beta$ model and a NFW-like
cusp model.
Following the suggestion of Ikebe et al. (1996) and Matsushita et
al. (1998) together, it can be said that 
X-ray bright galaxy groups and X-ray bright elliptical galaxies consist
of the same class of objects, which exhibit a double-$\beta$ model
structure.

\begin{figure}[hptb]
\begin{minipage}[tbhn]{8cm}
\centerline{\includegraphics[width=9.5cm]{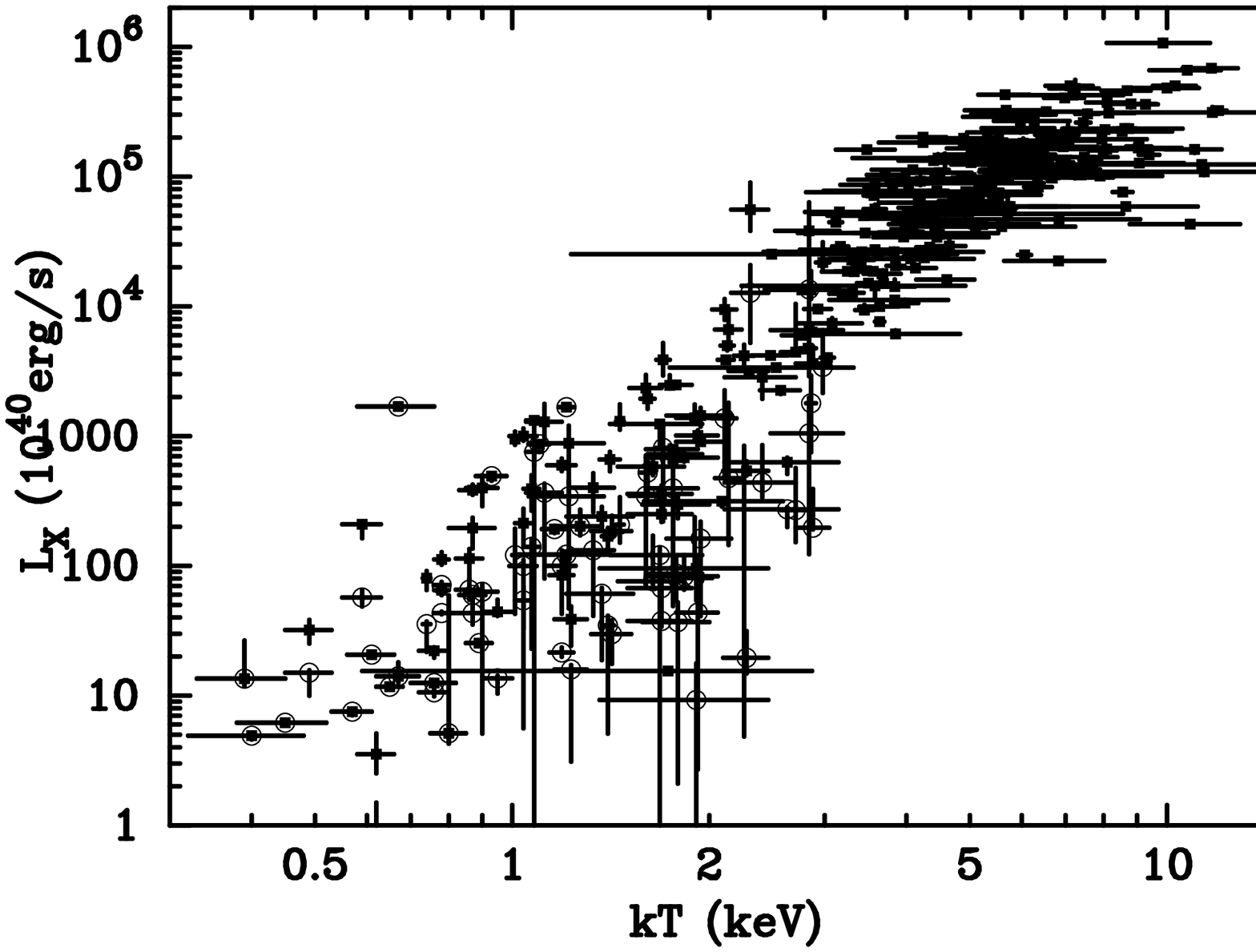}}
\caption{Bolometric luminosity against the gas temperature. The
 filled squares are the same as in figure \ref{tlx}, and the open circles
 represent the luminosity of the inner component in the double-$\beta$
 model fitting.}
\label{tlx2}
\end{minipage}\quad
\begin{minipage}[tbhn]{8cm}
\centerline{\includegraphics[width=9.5cm]{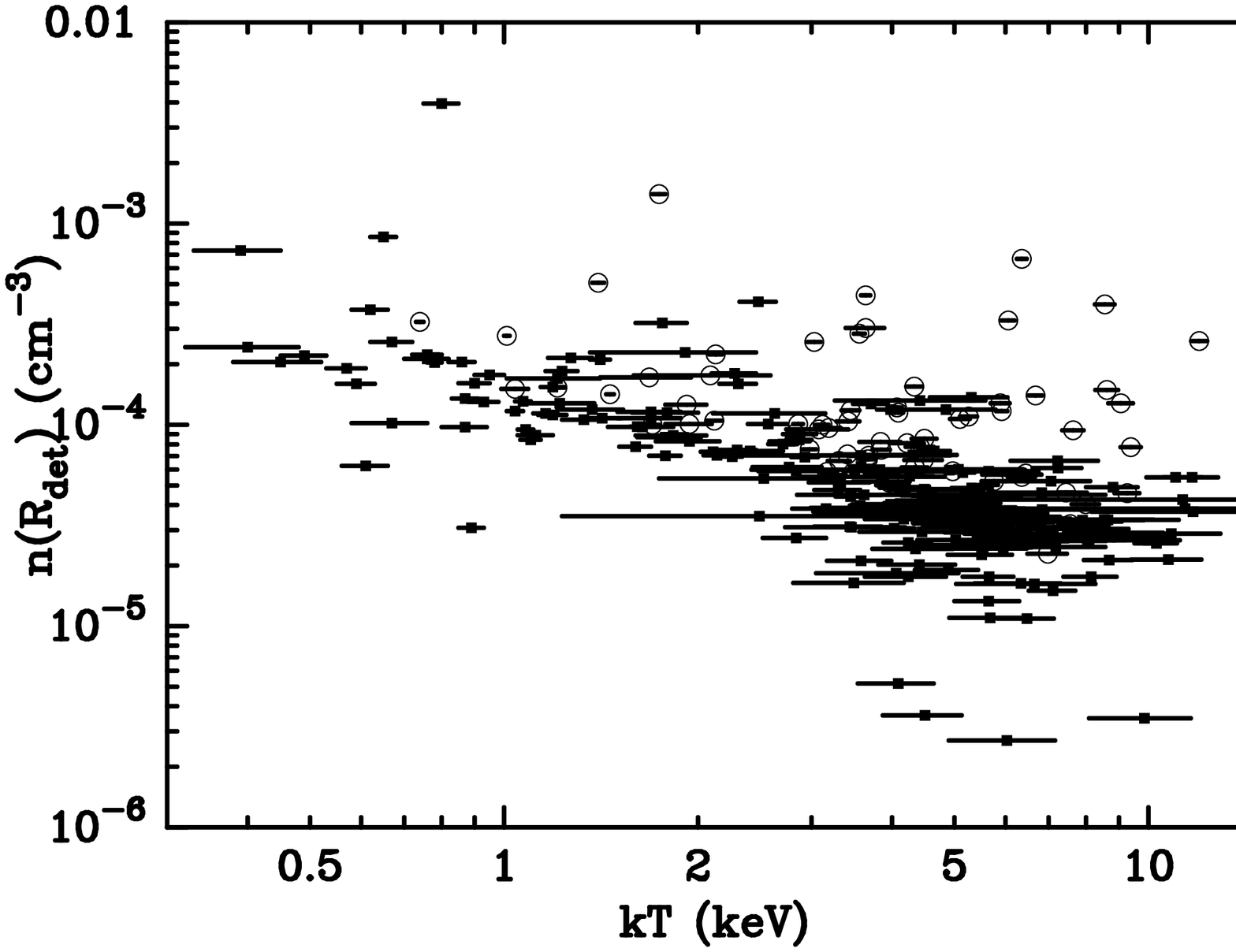}}
\caption{Electron density at the maximum detection radius, 
$R_{\rm limit}$. The open circles represent the case where the emission
 is beyond the GIS field of view.}
\label{t-nrdet}
\end{minipage}
\end{figure}

In figure \ref{tlx2}, we plot the luminosity of the inner $\beta$-model
component against the temperature, together with the total luminosity.
It can be seen that the inner components and X-ray faint elliptical
galaxies connect smoothly with each other, indicating that we might see
the same type of X-ray hot gas.
Accordingly, as described in subsection 4.4, two types
of X-ray hot gas come out;
two components of the double-$\beta$ model are
associated with the potentials of the galaxy and cluster.
The cluster component is dominant in rich clusters, and not seen in
X-ray faint elliptical galaxies.
The galaxy component is seen in elliptical galaxies and poor clusters,
and not seen in rich clusters.
The reason for this phenomenon might be due to the temperature
dependence of the X-ray distribution of the cluster component.
We have already showed that the central electron density of the outer component in
the double-$\beta$ model is lower for lower-temperature objects.
In other words, the X-ray surface brightness of the outer component 
becomes fainter for lower-temperature objects.
Therefore, in lower-temperature objects, the cluster component cannot be
detected and we can only observe the galaxy component, while
in rich clusters, where hot gas of the cluster component is
bright and the gravitational potential of the galaxy is relatively
negligible, the hot gas of the galaxy component cannot be resolved.
Sanderson et al. (2003) claimed that the properties of the X-ray halo 
of elliptical galaxies in their sample are different from those of
galaxy clusters, and we also confirmed this trend for a larger sample.
They attributed this difference to the earlier formation epoch than that
of rich clusters.
However, we suggest that the X-ray size of galaxy-scale
hot gas is intrinsically compact due to the smaller scale of galaxy
potential than that of the cluster-scale one.

Therefore, it is speculated that, 
even in X-ray faint elliptical galaxies, 
the group-scale hot gas exists, but it is too faint to detect.
If this hypothesis is correct, we can claim the general view of
elliptical galaxies; 
most elliptical galaxies associate the group-scale
gravitational potential and thus hot gas, but the density of hot
gas scatters widely.
Figure \ref{t-nrdet} supports the above suggestion, 
indicating that, for objects with $kT<$1 keV, 
the hot-gas density at the maximum detection
radius in the GIS data is higher than $10^{-4}$ cm$^{-3}$, and
different by only several times from the central hot-gas density (figure
\ref{betafit}).
In other words, the X-ray surface brightness of extended group-scale hot
gas for lower-temperature objects is around the detection
threshold, and sometimes cannot be detected even if it exists.
In this case, we cannot distinguish whether the extended group-scale hot
gas exists or not by the present data,
and Astro-E2 XIS with large effective area and low background level will
provide an opportunity to do it.

Our sample lacks X-ray faint galaxy groups, and HCG 68 is the only
object.
The X-ray surface brightness of HCG 68 is centered on the elliptical
galaxy NGC 5353, and can be fitted by a single-$\beta$
model. 
The maximam radius is at most 70$h_{50}^{-1}$ kpc, similar to
that of X-ray faint elliptical galaxies.
Therefore, at least the X-ray emission is not due to the group scale, but
due to the galaxy scale for HCG 68. 
In ROSAT PSPC observations, the extent was reported to be
$\sim130h_{50}^{-1}$ kpc (Pildis et al. 1995), and thus group-scale
X-ray hot gas is indicated.
In any case, the X-ray emission of HCG 68 seems to be dominated by the galaxy
component.
Mulchaey (2000) indicated that, for the lower temperature galaxy groups,
the X-ray emission becomes irregular and is dominated by individual member 
galaxies.
The opposite cases are spiral-dominant galaxy groups, such as HCG 92
(Stephan's Quintet) (Sulentic et al. 1995; Awaki et al. 1997), HCG 57
(Fukazawa et al. 2002), and HCG 16 (Besole et al. 2003).
Since these systems do not contain a dominant elliptical galaxy, the
X-ray hot gas of the galaxy-scale component is not bright.
Alternatively, the X-ray emission is dominated by the faint diffuse
component with a size of 100--200$h_{50}^{-1}$ kpc.
This is naturally considered to be the group-scale hot gas.
In summary, there are two types of X-ray faint objects.
One is dominated by the elliptical galaxy in the X-ray emission, and the
group-scale hot gas is hardly, or barely, confirmed.
The other is a spiral-dominated system, where we can observe 
only the group-scale hot gas with very faint diffuse X-ray emission.

We suggest that X-ray faint galaxy groups and elliptical galaxies may
contain an amount of hot gas that is as large as X-ray bright ones, and there
would be a non-detected hot gas around them.
How can we observe them?
This depends on the temperature of the hot gas.
In our scenario,
such hot gas would be heated up by gravity as well as hot gas at the 
inner region; therefore, the temperature is thought to be 0.5--1 keV.
In this case, the detection of thermal X-ray emission is the most
possible case,
but is quite difficult.
In the ASCA data, the systematic uncertainty of the CXB intensity rather
than the detector intrinsic background limits the
sensitivity.
Both a wide field of view and good imaging quality to resolve CXB into point
sources are necessary, and such detectors are not presently being 
proposed for the future mission.
The other situation is that the hidden hot-gas component is not
virialized well, and the temperature is around $10^{4-6} K$.
The detection of such a thermal emission is difficult, and
absorption lines in the spectra of background objects are expected.
This indication is claimed by Mulchaey et al. (1996b), and so on.
However, we must explain why the hot gas in the outer region is not
thermalized.
As described in subsection 5.1, we showed that the small detection radius and low
luminosity of X-ray faint systems are not due to the compactness of the 
hot-gas extent, but due to the detection sensitivity.
Several numerical simulations show that the ICM is heated up by shock
waves within a radius of $r_{500}$ (Evrard et al. 1996; 
Takizawa, Mineshige 1998).
Therefore, it is reasonable to think that the hot-gas temperature is not
so different from the inner region as objects whose detection radius is
large.

\subsection{Scatter of X-ray Luminosity of Galaxy Groups and Elliptical Galaxies}

Ao far we suggested, two hot-gas components exist in lower temperature systems:
a galaxy component and a group component.
Considering these components, we discuss the scatter of the X-ray properties
for elliptical galaxies and galaxy groups.
As mentioned in subsection 5.1, 
we imply that these two systems are basically the
same class of objects for hot gas and dark matter, and only the stellar
distribution is different; stars in elliptical galaxies concentrate in
one galaxy, while stars in galaxy groups separately locate in several
galaxies.
This difference may be due to whether the galaxies in the systems 
have merged or not.
The scatter of the X-ray luminosity is mainly caused by the scatter of
the hot-gas density for the group components and the brightness of the 
galaxy components.
The origin of the former is related to the degree of heating, dark-matter
concentration, system age, and so on, and may be explained in 
formation theories.
On the other hand, the scatter of the X-ray brightness of the 
galaxy components is not simply understood.
Matsushita (1997, 2001) showed that the hot-gas mass within 
the 4-times effective
radius is systematically different between X-ray bright ones and X-ray faint
ones.
Therefore, the difference at such a small scale is thought to be attributed 
not to the group component, but to the cD elliptical galaxies.
The compression of the galaxy hot-gas component by the 
high pressure of the group hot-gas component in X-ray bright ones, the 
escape of hot gas from X-ray faint
ones, and so on, are considered, although we cannot investigate this
issue with the ASCA data.
High-resolution imaging with Chandra data will give an answer.

The authors thank an anonymous referee for a careful reading and 
many helpful comments.
The authors acknowledge T. Takahashi and the ASCA image analysis working
group (Takahashi et al. 1995).
The authors are also grateful to the ASCA team for
their help in the spacecraft operation and calibration.
YF thanks Prof. T. Ohsugi for encouraging and supporting this study.

\setcounter{figure}{12}
\begin{onecolumn}
\begin{figure}[hptb]
\begin{minipage}[tbhn]{8cm}
\centerline{\includegraphics[width=9.5cm]{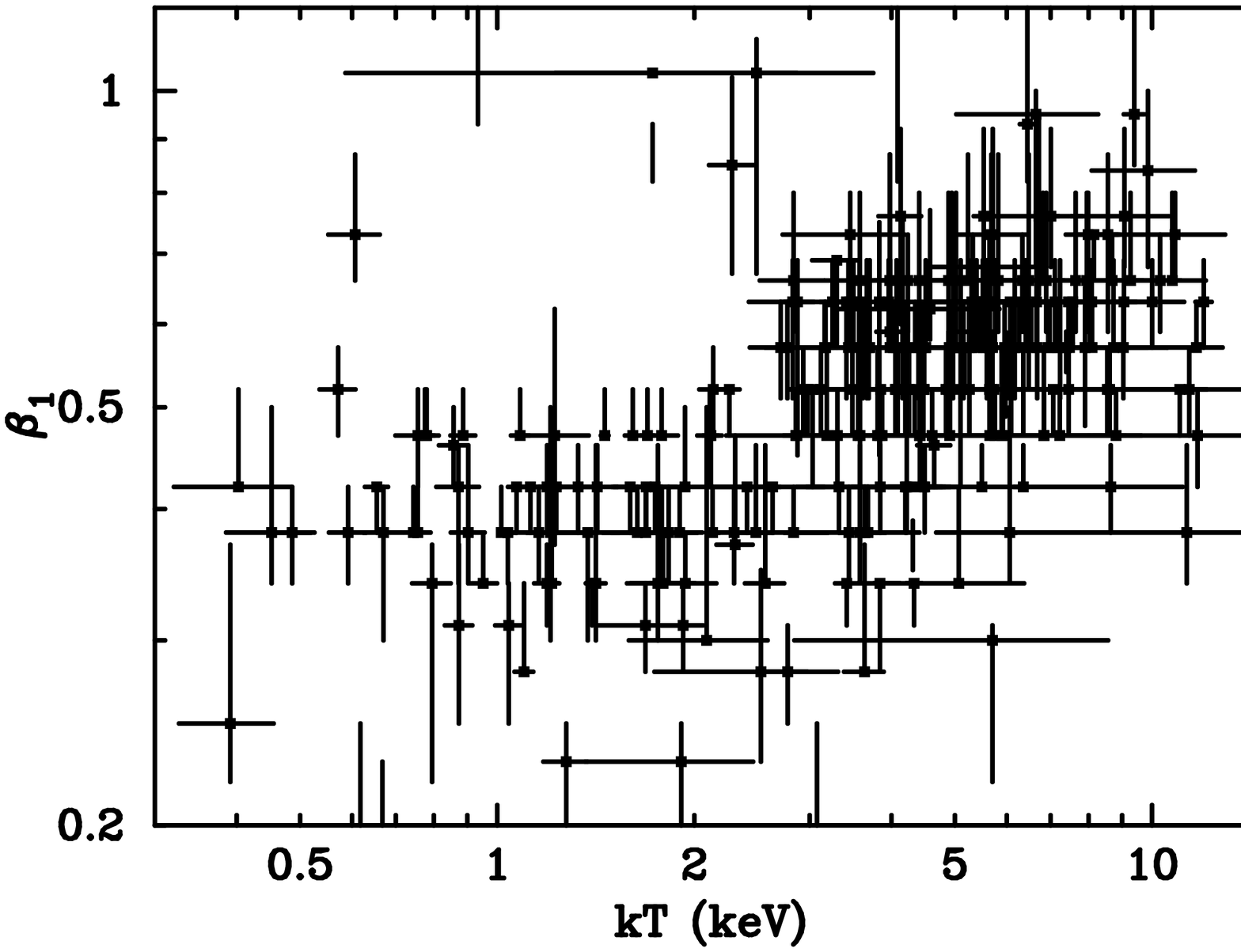}}
\end{minipage}\quad
\begin{minipage}[tbhn]{8cm}
\centerline{\includegraphics[width=9.5cm]{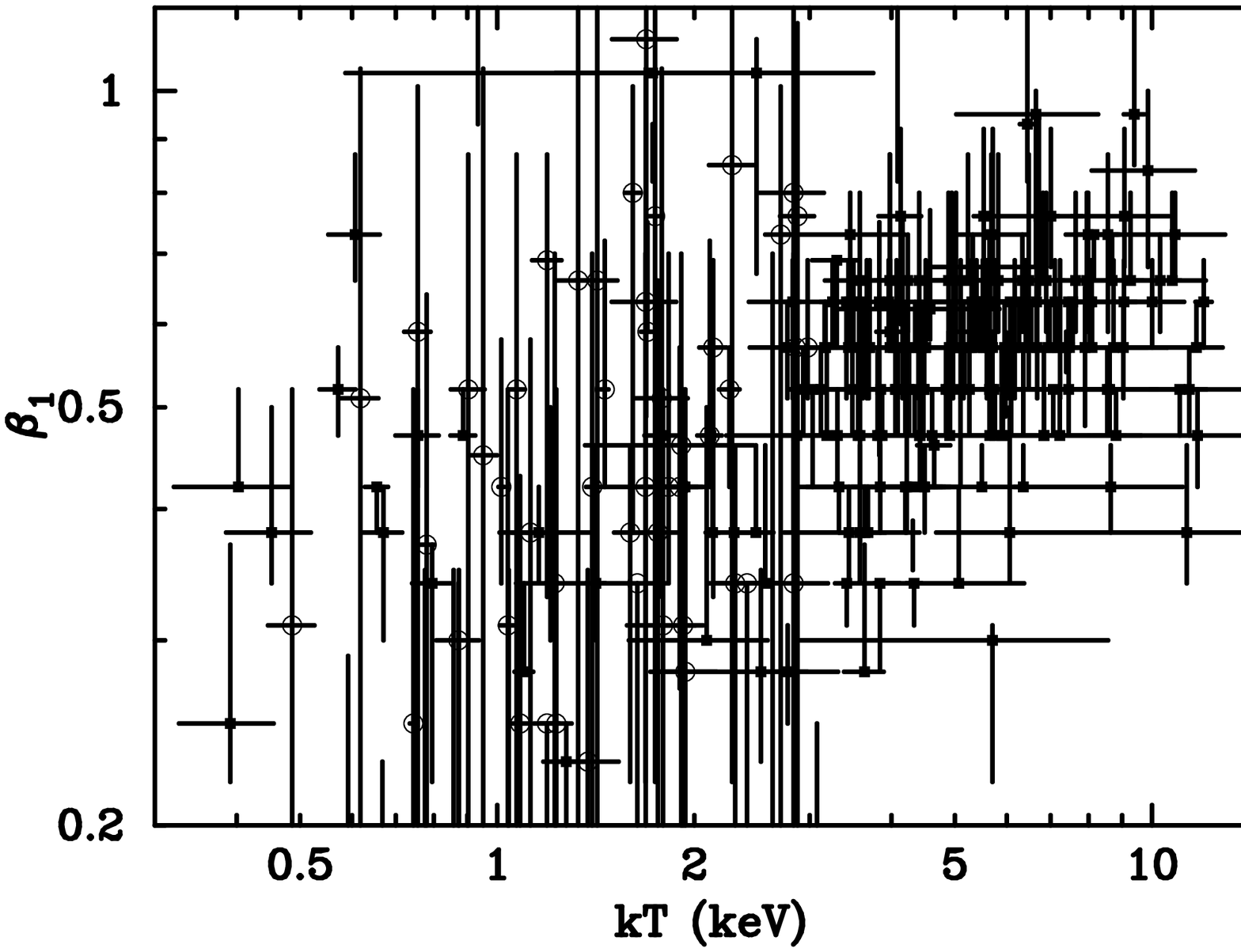}}
\end{minipage}
\begin{minipage}[tbhn]{8cm}
\centerline{\includegraphics[width=9.5cm]{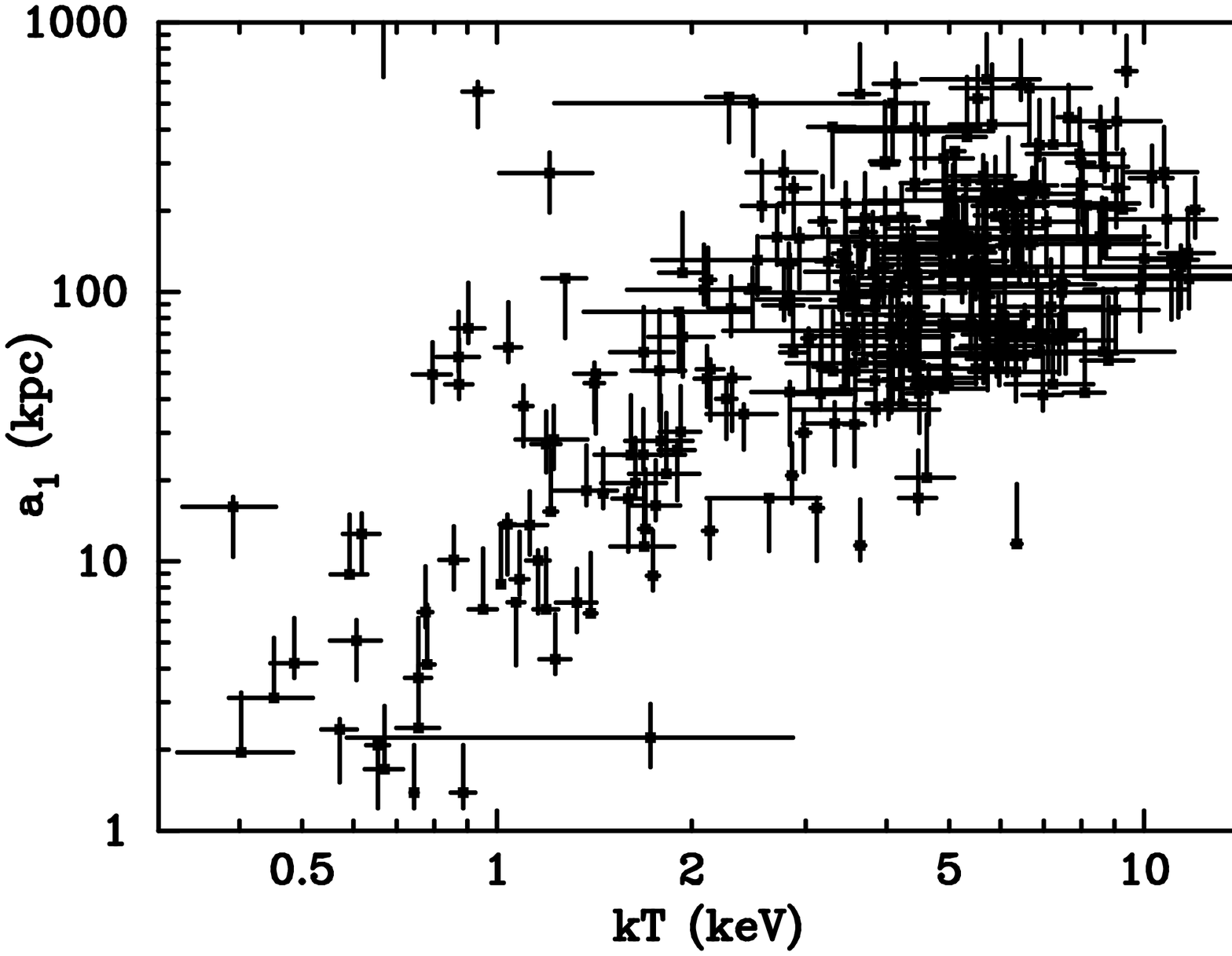}}
\end{minipage}\quad
\begin{minipage}[tbhn]{8cm}
\centerline{\includegraphics[width=9.5cm]{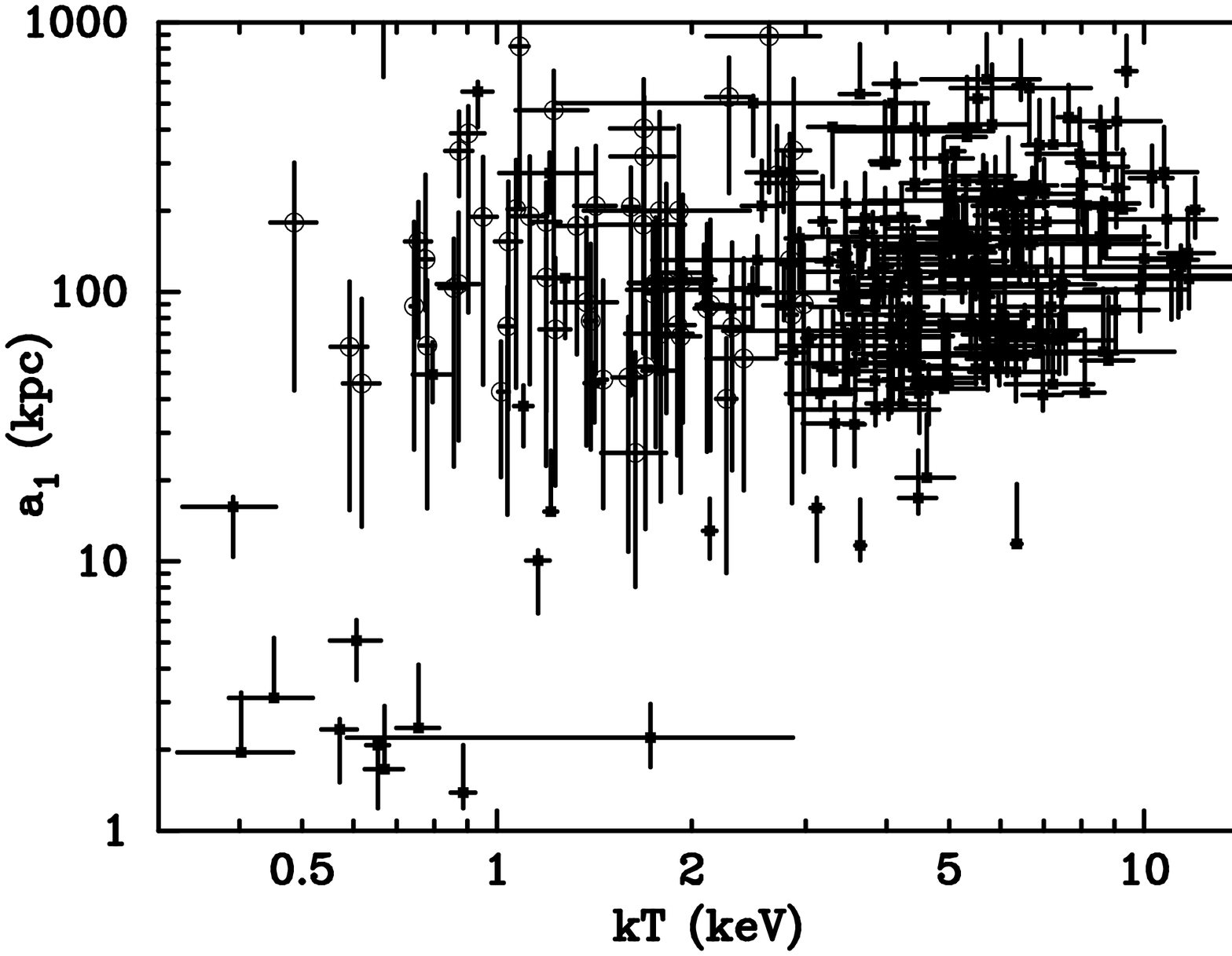}}
\end{minipage}
\begin{minipage}[tbhn]{8cm}
\centerline{\includegraphics[width=9.5cm]{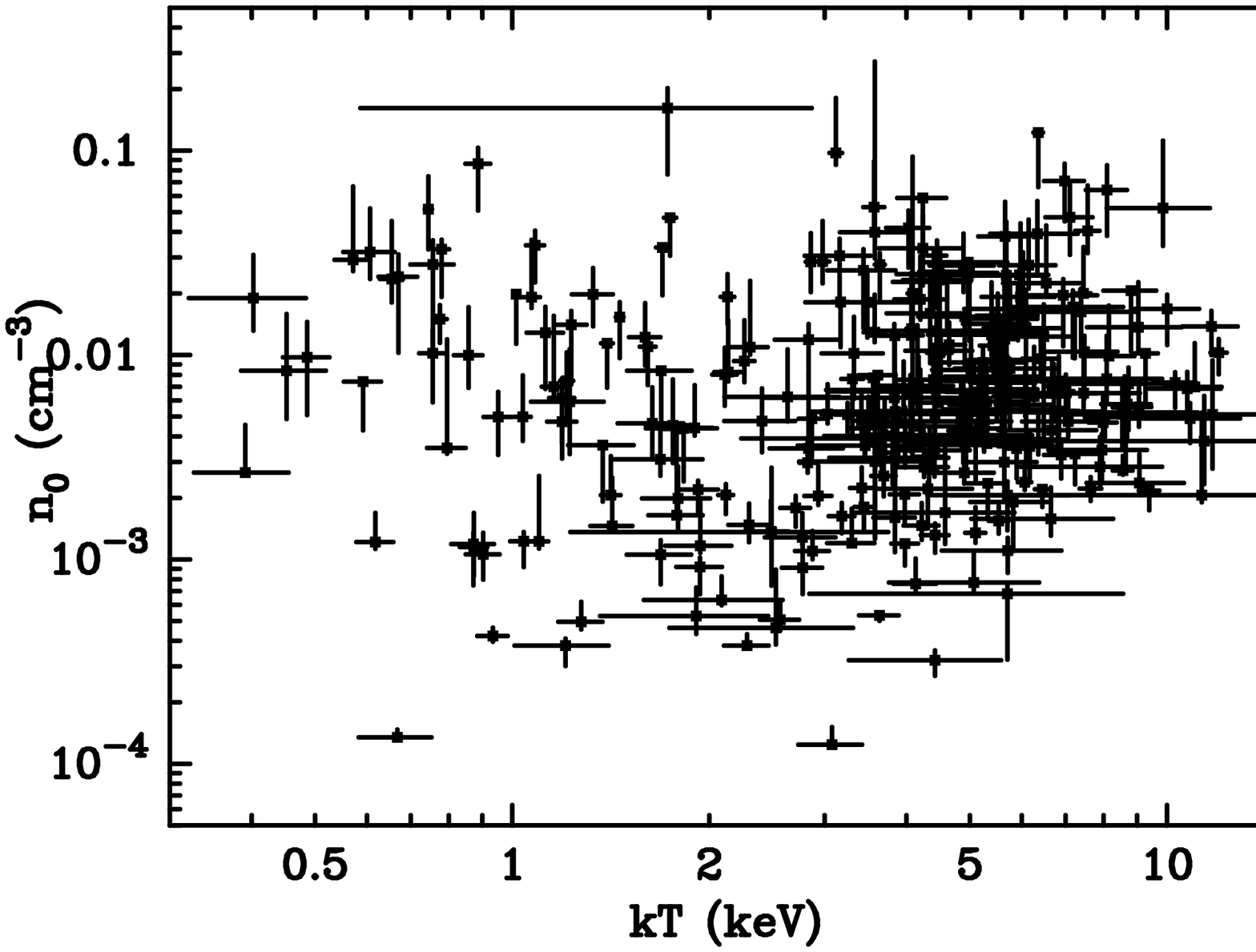}}
\end{minipage}\quad
\begin{minipage}[tbhn]{8cm}
\centerline{\includegraphics[width=9.5cm]{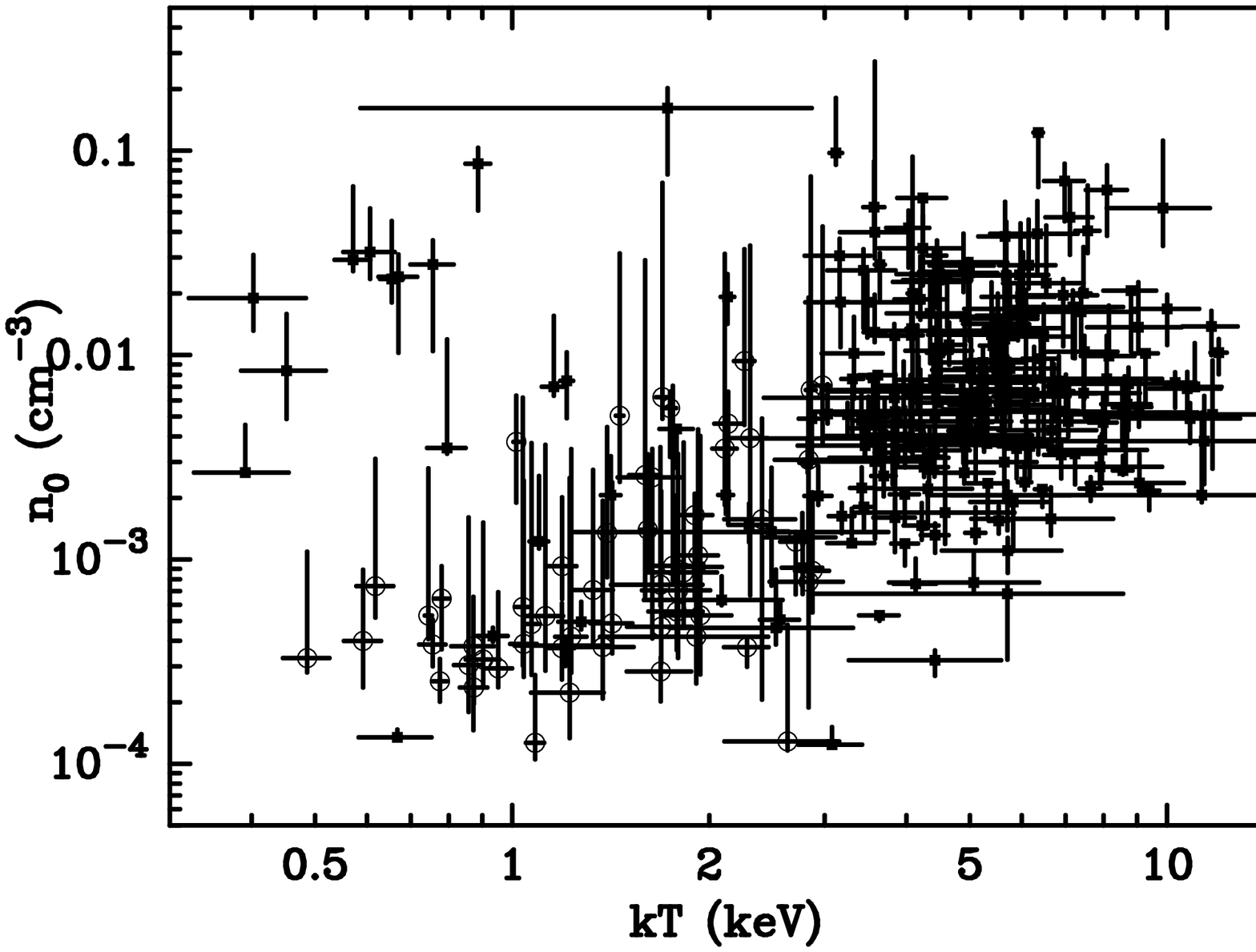}}
\end{minipage}
\caption{Gas-temperature dependence of the $\beta$-model
 parameters. The top, middle, and bottom panels show $\beta$, the core
 radius, and the central electron density, respectively. The left panels 
 plot the results of single-$\beta$ model fittings. The right panels
 plot the results of double-$\beta$ model fittings for the objects listed in
 table \ref{b2fit-lst} (open circle), 
 together with those of single-$\beta$ model
 fittings for other objects (filled square).}
\end{figure}
\end{onecolumn}
\clearpage

\clearpage

\footnotesize


\begin{center}
\begin{table}
\caption[]{Sample list of clusters of galaxies.}
\label{sample-lst}
\caption[]{Results of spectral fittings.}
\label{specfit-lst}
\caption[]{Results of radial profile fittings.}
\label{b1fit-lst}
\caption[]{Flux, central surface brightness, and luminosity.}
\label{fxlx-lst}
\caption[]{Results of radial profile fittings with double-$\beta$ model.}
\label{b2fit-lst}
\end{table}
\end{center}


\end{document}